\pdfoutput=1

\documentclass[11pt,nofootinbib]{article}
\usepackage{amsmath,amssymb,mathtools}
\usepackage{graphicx}
\usepackage[colorlinks=true,linkcolor=blue,citecolor=blue]{hyperref}
\usepackage{subfigure}
\usepackage{comment}

\numberwithin{equation}{section}

\begin{document}

\providecommand{\abs}[1]{\lvert#1\rvert}

\begin{titlepage}

\setcounter{page}{1} \baselineskip=15.5pt \thispagestyle{empty}

\begin{flushright}
\end{flushright}
\vfil


\bigskip
\begin{center}
 {\LARGE \textbf{Primordial Magnetic Fields}}\\
\medskip 
 {\LARGE \textbf{from the Post-Inflationary Universe}}
\vskip 15pt
\end{center}

\vspace{0.5cm}
\begin{center}
{\Large 
Takeshi Kobayashi
}
\end{center}

\vspace{0.3cm}

\begin{center}
\textit{Canadian Institute for Theoretical Astrophysics,
 University of Toronto, \\ 60 St. George Street, Toronto, Ontario M5S
 3H8, Canada}\\ 

\vskip 14pt
\textit{Perimeter Institute for Theoretical Physics, \\ 
 31 Caroline Street North, Waterloo, Ontario N2L 2Y5, Canada}\\

\vskip 14pt
\textit{E-mail: takeshi@cita.utoronto.ca}

\end{center} 



\vspace{1cm}

\noindent
We explore cosmological magnetogenesis in the post-inflationary
 universe, when the inflaton oscillates around its potential minimum and
 the universe is effectively dominated by cold matter. During this epoch
 prior to reheating, large-scale magnetic fields can be significantly
 produced by the cosmological background. By considering
 magnetogenesis both during and after inflation, we demonstrate that
 magnetic fields stronger than $10^{-15}\, \mathrm{G}$ can be generated
 on Mpc scales without having strong couplings in the theory, or
 producing too large electric fields that would dominate the
 universe.

\vfil

\end{titlepage}

\newpage
\setcounter{tocdepth}{2}
\tableofcontents

\section{Introduction}
\label{sec:intro}

Large-scale magnetic fields exist in cosmic structures such as galaxies
and galaxy clusters, however the origin of the fields remain
unexplained. Recent gamma ray observations suggest the existence of
magnetic fields even in void
regions~\cite{Tavecchio:2010mk,Neronov:1900zz,Ando:2010rb,Taylor:2011bn,Takahashi:2013uoa,Finke:2013bua}, 
and the amplitude of such intergalactic magnetic fields were derived to
be stronger than $\sim 10^{-15}$G. 
Although this lower bound has astrophysical uncertainties (see
e.g.~\cite{Broderick:2011av,Miniati:2012ge}), such large-scale magnetic
fields hint that the magnetic fields are of cosmological origin. 

Here it should be noted that the equations of motion of the electromagnetic
fields in a Friedmann universe reduce to those in Minkowski space,
since the standard Maxwell theory is conformally invariant.
Therefore the conformal invariance should be broken for the
electromagnetic fields to be significantly produced by the cosmological
background. Such mechanisms of cosmological magnetogenesis have been
embedded in the inflationary universe in various works, 
e.g.,~\cite{Turner:1987bw,Ratra:1991bn,Garretson:1992vt,Gasperini:1995dh,Giovannini:2000dj,Davis:2000zp,Bamba:2003av,Martin:2007ue,Demozzi:2009fu,Kanno:2009ei}.\footnote{As
an alternative way of cosmologically producing magnetic fields,
\cite{Vachaspati:1991nm,Cornwall:1997ms,Vachaspati:2008pi} have studied
magnetogenesis during phase transitions.} 
However the studies have revealed that inflationary magnetogenesis
cannot produce significant magnetic fields without running into at
least one of the inconsistencies:
(i) the electromagnetic fields obtain strong couplings and therefore the
theory becomes
uncontrollable~\cite{Gasperini:1995dh,Demozzi:2009fu,Fujita:2012rb},
(ii) too large electric fields are produced such that their backreaction can 
spoil inflation and/or
magnetogenesis~\cite{Bamba:2003av,Demozzi:2009fu,Kanno:2009ei,Fujita:2012rb}. 
In particular, the backreaction of the electric fields is a problem
inherent to inflationary magnetogenesis;
after magnetogenesis the magnetic fields decay as $B \propto a^{-2}$
where $a$~is the scale factor,
therefore the present day magnetic fields should be
remnants of substantial photon production 
during a short period of time in the very early universe.
In other words, the electromagnetic vector potential during
inflation is required to possess a large time-derivative, and therefore
leads to excessive production of electric fields.
Recently it has also been pointed out that the generated electromagnetic
fields can induce cosmological density perturbations beyond the observed
value, and
works~\cite{Suyama:2012wh,Giovannini:2013rme,Ringeval:2013hfa,Nurmi:2013gpa,Fujita:2014sna} 
have imposed further constraints on inflationary magnetogenesis. 
(See
also~\cite{Shaw:2010ea,Yamazaki:2012pg,Camera:2013fva,Berger:2014wta}
and references 
therein for constraints on primordial magnetic fields from CMB and
large-scale structure data.)
However, we stress that the inflationary expansion itself
is not a necessary condition for magnetogenesis; it is the breaking of
the conformal invariance that allows magnetic field production by the
cosmological background.

In this paper, we investigate post-inflationary magnetogenesis by
breaking the conformal invariance of the Maxwell theory after
inflation. 
The conductivity of the universe becomes high during
reheating and thereafter the magnetic fields decay
as $B \propto a^{-2}$~\cite{Turner:1987bw}. Therefore we focus on the
epoch between the end of inflation and (p)reheating, 
during which the inflaton field oscillates around its potential minimum and
the universe is effectively dominated by cold matter. 
The conformal invariance of the Maxwell theory is broken by couplings
between the electromagnetic fields and scalar degrees of freedom which
can be the inflaton or some other spectator field(s).
By considering magnetogenesis both during and after inflation, we 
demonstrate that magnetic fields of~$10^{-15}$G or stronger can be
produced at cosmological scales (say, on Mpc scales), without running
into the strong coupling regime or producing too much electric fields. 
The problem of affecting the cosmological density perturbations is also 
ameliorated since the magnetic fields are enhanced after inflation.
The proposed model is compatible even with high scale inflation.

This paper is organized as follows. We first review the electromagnetic
theory in an expanding universe in Section~\ref{sec:EMFRW}.
Then we move on to study magnetogenesis during the 
inflationary epoch in Section~\ref{sec:during}, and
post-inflationary magnetogenesis in 
Section~\ref{sec:after}. 
Here we will see that the magnetogenesis in each epoch alone
are highly constrained by the strong coupling and backreaction
problems. In particular, the constraints on inflationary and
post-inflationary magnetogenesis will be translated into severe upper
bounds on the inflation and reheating scales, respectively.
Then in Section~\ref{sec:model} we discuss the combined scenario of
inflationary {\it and} post-inflationary magnetogenesis.
There we will see that such a two step model can overcome the challenges
and efficiently produce large-scale magnetic fields. 
In Section~\ref{sec:couplings}, we present some examples of
scalar couplings that break the conformal invariance of the Maxwell
theory. Finally, we conclude in Section~\ref{sec:conc}.

\section{Electromagnetic Fields in an Expanding Universe}
\label{sec:EMFRW}

Let us start by reviewing the evolution of electromagnetic fields in an
expanding universe. For a more detailed review
see~\cite{Widrow:2002ud,Barrow:2006ch,Kandus:2010nw}. 

Throughout this paper we study an electromagnetic
theory with an effective coupling~$I$,
\begin{equation}
 S = \int d^4 x \sqrt{-g} 
\left(
-\frac{I^2}{4} 
F_{\mu \nu} F^{\mu \nu} 
\right),
\label{maxwell}
\end{equation}
where $F_{\mu \nu} = \partial_\mu A_\nu - \partial_\nu A_\mu$ and
the Greek indices take values $\mu, \nu = 0,1,2,3$.
The coupling~$I$ is an arbitrary function of other degrees of
freedom, for instance, of a scalar field,
$I = I(\sigma)$~\cite{Ratra:1991bn}. 
The standard Maxwell theory is recovered when $I$ is a constant, then
the theory is invariant under the conformal transformation $g_{\mu \nu}
\to \Omega^2 g_{\mu \nu}$.\footnote{Alternatively, the conformal
invariance of the electromagnetic fields can be broken by a photon mass
term $m_\gamma^2 A_\mu A^\mu$. However significant magnetogenesis
requires a tachyonic mass $m_\gamma^2 <0$, which can arise from,
e.g., non-minimal couplings between the electromagnetic field and
gravity~\cite{Turner:1987bw}. 
However such a theory contains several problems including the
existence of a ghost~\cite{Dvali:2007ks,Himmetoglu:2009qi}. (See
also discussions in~\cite{Demozzi:2009fu,Fujita:2012rb}.)
Therefore in this paper we focus on massless photons whose conformal
invariance is broken through the kinetic coupling~(\ref{maxwell}).}

It should also be noted that $I$ being tiny, i.e. $I^2 \ll 1$,
can lead to strong couplings in the electromagnetic theory.
Considering an interaction term with a charged fermion such as
$c \bar{\psi} \gamma^\mu \psi A_\mu$ where $c$ is a coupling constant,
then after canonically normalizing the field as $\tilde{A}_\mu = I A_\mu$, the 
interaction term becomes $\frac{c}{I}  \bar{\psi} \gamma^\mu \psi
\tilde{A}_\mu$ and one sees that the effective coupling is $c / I$. 
Thus a tiny~$I$ would push the theory into the strong
coupling regime where perturbative calculations break down. 

\subsection{Quantization of Electromagnetic Fields}

We consider a flat FRW background
\begin{equation}
 ds^2 = g_{\mu \nu} dx^\mu dx^\nu = 
a(\tau)^2 \left( -d\tau^2 + 
d \boldsymbol{x}^2 \right),
\label{flatFRW}
\end{equation}
where $\tau$ is the conformal time. 
The coupling~$I$ is also considered to be homogeneous, i.e., 
$I = I(\tau)$. 
(In case where $I$ is a function of a scalar field~$\sigma$, then 
we suppose $\sigma = \sigma(\tau)$.)
We decompose the spatial components of the vector potential~$A_i$ into
irrotational and incompressible parts,
\begin{equation}
 A_\mu = \left( A_0, \, \partial_i S  + V_i \right),
\end{equation}
where
\begin{equation}
 \partial_i V_i = 0. \label{transverse}
\end{equation}
The Latin letters denote spatial indices $i, j = 1,2,3$, and the sum over repeated 
spatial indices is implied irrespective of their positions.
Then the action~(\ref{maxwell}) can be rewritten as, up to total
derivatives, 
\begin{equation}
 S = \int d\tau d^3 x  \, \frac{I^2}{2} 
\left\{
V_i'V_i' + \partial_i S' \partial_i S' +
\partial_i A_0 \partial_i A_0
- 2 \partial_i S' \partial_i A_0
- \partial_i V_j \partial_i V_j
\right\},
\label{Smid}
\end{equation}
where a prime denotes a $\tau$-derivative. 
Varying the action in terms of the Lagrange multiplier~$A_0$
and choosing proper boundary conditions, we obtain
\begin{equation}
 A_0 = S'. \label{sixstar}
\end{equation}
By substituting this constraint into the action (\ref{Smid}), the scalar
modes~$A_0$ and $S$ vanish and we are left with the two polarization
states of the photon (note the constraint~(\ref{transverse})),
\begin{equation}
 S = \int d\tau d^3x \, \frac{I^2}{2}
\left\{ 
V_i' V_i' - \partial_i V_j \partial_i V_j
\right\}.
\label{S2.7}
\end{equation}
The equation of motion of~$V_i$ reads
\begin{equation}
 V_i'' + 2 \frac{I'}{I} V_i' - \partial^2 V_i = 0,
\label{ViEoM}
\end{equation}
where $\partial^2 \equiv \partial_j \partial_j$. 
Now considering the Fourier modes,
\begin{equation}
 V_i (\tau, \boldsymbol{x}) = 
\frac{1}{(2 \pi )^3} \int d^3 k\, 
e^{i \boldsymbol{k \cdot x}} \xi_i (\tau, \boldsymbol{k}),
\label{fourier}
\end{equation}
then $\xi_i k_i  = 0$ should be satisfied from the 
constraint~(\ref{transverse}). 
Let us express $\xi_i $ as a linear combination of two orthonormal
polarization vectors~$\epsilon_i^{(p)}(\boldsymbol{k})$ with $p = 1,2$, satisfying
\begin{equation}
 \epsilon_i^{(p)} (\boldsymbol{k}) \,  k_i = 0, 
\qquad
 \epsilon_i^{(p)} (\boldsymbol{k}) \epsilon_i^{(q)} (\boldsymbol{k})  =
 \delta_{pq}.
\label{2.10}
\end{equation}
Note that from (\ref{2.10}) follows
\begin{equation}
 \sum_{p = 1,2} \epsilon_i^{(p)} (\boldsymbol{k})
\epsilon_j^{(p)} (\boldsymbol{k})
 = \delta_{ij} - \frac{k_i k_j}{k^2},
\label{epepsum}
\end{equation}
where $k \equiv |\boldsymbol{k}|$. 
Unlike the spacetime indices, we do {\it not} assume implicit summation
over the polarization index~$(p)$ throughout this paper. 

Let us now quantize the theory by promoting $V_i$~(\ref{fourier}) to an operator,
\begin{equation}
 V_i(\tau, \boldsymbol{x}) = \frac{1}{(2 \pi)^3}
 \sum_{p = 1,2} \int d^3 k \, \epsilon^{(p)}_i (\boldsymbol{k})
\left\{
e^{i \boldsymbol{k \cdot x}}  a_{\boldsymbol{k}}^{(p)} 
u^{(p)}_{\boldsymbol{k}} (\tau) + 
e^{-i \boldsymbol{k \cdot x}} a_{\boldsymbol{k}}^{\dagger (p)}
u^{*(p)}_{\boldsymbol{k}} (\tau)  
\right\}.
\label{Viop}
\end{equation}
Here $a_{\boldsymbol{k}}^{(p)}$ and $a_{\boldsymbol{k}}^{\dagger (p)}$ 
are annihilation and creation operators, respectively, and
$u^{(p)}_{\boldsymbol{k}} (\tau) $ is a mode function that satisfies the 
equation of motion (cf.~(\ref{ViEoM})),
\begin{equation}
 u''^{(p)}_{\boldsymbol{k}}+ 2\frac{I'}{I} u'^{(p)}_{\boldsymbol{k}}
+ k^2  u^{(p)}_{\boldsymbol{k}} = 0.
\label{maru1}
\end{equation}
Here it should be noted that the time evolving coupling~$I$ gives rise to a
(positive or negative) friction term; the standard Maxwell theory with
$I = \mathrm{const.}$ brings the equation of motion to the wave equation in a
flat spacetime. 

The conjugate momentum of $V_i$ is obtained from the action 
$S = \int d\tau d^3x \, \mathcal{L}$ in (\ref{S2.7}) as
\begin{equation}
 \Pi^i = \frac{\partial \mathcal{L}}{\partial V_i'} = I^2 V_i',
\end{equation}
and the commutation relations are imposed as
\begin{equation}
 [ a_{\boldsymbol{k}}^{(p)},\,  a_{\boldsymbol{h}}^{(q)} ] =
 [ a_{\boldsymbol{k}}^{\dagger (p)},\,  a_{\boldsymbol{h}}^{\dagger (q)}
 ] = 0,
\qquad 
 [ a_{\boldsymbol{k}}^{(p)},\,  a_{\boldsymbol{h}}^{\dagger (q)} ] = (2
  \pi)^3 \, 
\delta^{pq} \,
\delta^{(3)}  (\boldsymbol{k} - \boldsymbol{h}) , 
 \label{eq:commu}
\end{equation}
\begin{equation}
 \left[ V_i(\tau, \boldsymbol{x}),\,  V_j (\tau, \boldsymbol{y}) \right] = 
 \left[ \Pi^i(\tau, \boldsymbol{x}),\,  \Pi^j (\tau, \boldsymbol{y}) \right]
 = 0,
\label{eq:commu2}
\end{equation}
\begin{equation}
 \left[ V_i(\tau, \boldsymbol{x}),\,  \Pi^j (\tau, \boldsymbol{y})
 \right] = i 
\delta^{(3)}   (\boldsymbol{x} - \boldsymbol{y})
\left( \delta_{ij} - \frac{\partial_i \partial_j}{\partial^2} \right).
\label{eq:commu3}
\end{equation}
Using (\ref{epepsum}), the relation (\ref{eq:commu3}) can be rewritten as
\begin{equation}
 \left[ V_i(\tau, \boldsymbol{x}),\,  \Pi^j (\tau, \boldsymbol{y})
 \right] = 
 \frac{i}{(2\pi)^3}\sum_{p = 1,2}\int d^3 k \, 
 e^{i\boldsymbol{k\cdot}  (\boldsymbol{x - y})}
\epsilon_i^{(p)} (\boldsymbol{k}) \epsilon_j^{(p)} (\boldsymbol{k}).
\label{eq:commu4}
\end{equation}
Choosing the polarization vectors such that $\epsilon^{(p)}_i
(\boldsymbol{k}) = \epsilon^{(p)}_i (-\boldsymbol{k})$, then one can check
that the commutation relations (\ref{eq:commu2}) and (\ref{eq:commu4})
follow from~(\ref{eq:commu}) when the mode function 
is independent of the direction of~$\boldsymbol{k}$, i.e., 
\begin{equation}
 u_{\boldsymbol{k}}^{(p)} = u_k^{(p)},
\label{maru3}
\end{equation}
and obeys the normalization condition
\begin{equation}
 I^2 \left(
u_k^{(p)} u_k^{'*(p)} - u_k^{*(p)} u_k^{'(p)}
\right) = i.
\label{maru2}
\end{equation}

\subsection{Power Spectra of Magnetic and Electric Fields}

The magnetic and electric fields measured by a comoving observer with
4-velocity~$u^\mu$ ($u^i=0$, $u_\mu u^\mu = -1$) is 
\begin{equation}
 B_\mu = \frac{1}{2} \varepsilon_{\mu \nu \sigma} F^{\nu \sigma},
\qquad
E_\mu = u^\nu F_{\mu \nu},
\end{equation}
where 
\begin{equation}
 \varepsilon_{\mu \nu \sigma} = \eta_{\mu \nu \sigma \lambda} u^\lambda,
\end{equation}
and $\eta_{\mu \nu \sigma \lambda}$ is  a totally antisymmetric
permutation tensor with $\eta_{0123} = -\sqrt{-g}$.
The time-components $B_0$ and $E_0$ vanish, and we find
\begin{equation}
 B_i = \frac{1}{a^4} \varepsilon_{ijk} \partial_j A_k,
\qquad
E_i = -\frac{1}{a} V_i'.
\end{equation}
Note that $\varepsilon_{ijk}$ is totally antisymmetric with
$\varepsilon_{123} = a^3$.
Hence the magnitude of the fields are given by
\begin{gather}
 B^2 \equiv B_\mu B^\mu = \frac{1}{a^2}B_i B_i = \frac{1}{a^4}
\left(\partial_i V_j \partial_i V_j - \partial_i V_j \partial_j V_i
\right),
\\
 E^2 \equiv E_\mu E^\mu = \frac{1}{a^2} E_i E_i  = \frac{1}{a^4}V_i' V_i' .
\end{gather} 
Choosing the vacuum as $ a_{\boldsymbol{k}}^{(p)} |0 \rangle = 0 $
for $p = 1,2$ and $^{\forall}  \boldsymbol{k}$,
then from (\ref{Viop}) and the commutation relation~(\ref{eq:commu}),
we obtain the correlation functions
\begin{align}
 \langle B_\mu (\tau, \boldsymbol{x}) B^\mu (\tau, \boldsymbol{y}) \rangle &=
 \int \frac{d^3 k}{4 \pi k^3}  e^{i\boldsymbol{k\cdot}  (\boldsymbol{x
- y})} \mathcal{P}_B (\tau, k),
\\
 \langle E_\mu (\tau, \boldsymbol{x}) E^\mu (\tau, \boldsymbol{y}) \rangle &=
 \int \frac{d^3 k}{4 \pi k^3}  e^{i\boldsymbol{k\cdot}  (\boldsymbol{x
- y})} \mathcal{P}_E (\tau, k),
\end{align}
where the power spectra are expressed in terms of the mode functions as
\begin{align}
 \mathcal{P}_B (\tau, k) &= \frac{k^5}{2 \pi^2 a(\tau)^4} 
\sum_{p=1,2} | u_k^{(p)} (\tau) |^2,
\label{PofB}
\\
 \mathcal{P}_E (\tau, k) &= \frac{k^3}{2 \pi^2 a(\tau)^4} 
\sum_{p=1,2} | u_k^{'(p)} (\tau) |^2.
\label{PofE}
\end{align}

Furthermore, the energy-momentum tensor of the electromagnetic fields is
\begin{equation}
 T^{\mathrm{EM}}_{\mu \nu} = I^2
\left( F_{\mu \sigma } F_\nu^{\sigma} - \frac{1}{4} 
g_{\mu \nu} F_{\sigma \lambda} F^{\sigma \lambda }
 \right),
\end{equation}
and the comoving observer measures the energy density as
\begin{equation}
 T^{\mathrm{EM} }_{\mu \nu} u^\mu u^\nu  
= - T_0^0 
= \frac{I^2}{2} \left(B^2 + E^2\right) .
\end{equation}
Hence the energy density of the electromagnetic fields are
expressed in terms of the power spectra (\ref{PofB}) and (\ref{PofE}) as
\begin{equation}
 \rho_{\mathrm{EM}}= 
\langle -T_0^0 (\tau, \boldsymbol{x})  \rangle =
\frac{I^2}{2} \int \frac{dk}{k} 
\left(  \mathcal{P}_B +  \mathcal{P}_E \right),
\label{rho_EM}
\end{equation}
which clearly shows the contribution from each of the magnetic and
electric fields.

\subsection{Evolution of Mode Functions}

Let us now see how the mode function~$u_k^{(p)}$ evolves as the universe
expands. We suppose a flat FRW background~(\ref{flatFRW}) with 
an equation of state parameter that takes a constant value except for~$-1/3$,
\begin{equation}
\frac{p_{\mathrm{bg}}}{\rho_{\mathrm{bg}} }= w = \mathrm{const.} \neq -\frac{1}{3}.
\end{equation}
Then the conformal time is expressed as
\begin{equation}
 d \tau = \frac{2}{1 + 3w} d \left( \frac{1}{aH}\right)
\label{conftime}
\end{equation}
in terms of the Hubble parameter $H = a'/a^2 = \dot{a}/a $, where an
overdot denotes a derivative with respect to the cosmological time~$dt =
a d\tau$. 

Here we consider a coupling~$I$ that scales as a power-law of the
scale factor, i.e.,
\begin{equation}
 I \propto a^{-s},
\label{Ipower-law}
\end{equation}
with a constant~$s$. Then the electromagnetic fields experience 
an effective ``horizon'' with radius of $\sim |I/\dot{I}| =
(|s|H)^{-1}$ (see the equation of motion~(\ref{maru1})), which 
is similar in size to the Hubble horizon up to the factor~$s$. 
We mainly focus on this case in the following
sections, and explicit realizations of such scaling behaviors through couplings with
scalar fields are discussed in Section~\ref{sec:couplings}.

Then the general solution for the mode function~$u_k^{(p)}$ in~(\ref{maru1}) is
given by Hankel functions multiplied by powers of~$aH$,
\begin{equation}
 (aH)^{-\alpha} H_\alpha^{(1),  (2)} 
\left(  \frac{2}{|1+3 w|}\frac{k}{aH} \right)
\qquad
\mathrm{with}
\qquad
\alpha = \frac{1}{2} + \frac{2s}{1 + 3 w}.
\label{Hankelsol}
\end{equation}

In the sub-horizon limit, i.e. $k/aH \to \infty$, 
the solutions~(\ref{Hankelsol}) are approximated by
\begin{multline}
\qquad 
 (aH)^{-\alpha} H_\alpha^{(1)} 
\left(  \frac{2}{|1+3 w|}\frac{k}{aH} \right)
  \\
 \simeq (aH)^{-\alpha}  
\left(\frac{|1+3 w|}{\pi } \frac{aH}{k} \right)^{1/2}
\exp \left\{
i\left(
\frac{2}{|1+3 w|}\frac{k}{aH} - \frac{ 2 \alpha +1}{4} \pi 
\right)
\right\} ,
\qquad
\label{subhlimit}
\end{multline}
and its complex conjugate for $(aH)^{-\alpha} H_\alpha^{(2)}$.

On the other hand, in the super-horizon limit, i.e. $k/aH \to 0$, each
of the two solutions in~(\ref{Hankelsol}) asymptote to a 
linear combination of the following terms:
\begin{equation}
\begin{split}
 (a H)^{-2 \alpha} ,
\quad
\mathrm{const.}
\qquad
\mathrm{for}
\quad
\alpha &\neq 0,
\\
 \ln \left( aH \right),
\quad
\mathrm{const.}
\qquad
\mathrm{for}
\quad
\alpha &= 0.
\label{superhlimit}
\end{split}
\end{equation}
In other words, $u_k^{(p)}$ outside the horizon consists of a
time dependent mode and a constant mode. 
These sets of super-horizon solutions can also be obtained directly from 
the equation of motion~(\ref{maru1}) by omitting the $k^2 u_k^{(p)}$
term, which gives the solutions 
\begin{equation}
 \int^\tau  \frac{d\tau}{I^2} ,
\quad
\mathrm{const.}
\end{equation}

\vspace{\baselineskip}

The standard Maxwell theory corresponds to the case of $s=0$,
i.e. $\alpha = 1/2$. Then the equation of motion~(\ref{maru1}) 
reduces to that in Minkowski space, and $u_k^{(p)}$ is a sum of plane
waves~$ e^{\pm i k \tau}$.\footnote{The super-horizon
solution~(\ref{superhlimit}) with $\alpha =  1/2$ 
holds a growing or decaying mode~$ \propto (aH)^{-1}$, in addition to the
constant solution.
This can also be seen by expanding the plane waves up to linear order
in~$k\tau$ as $\sin(k \tau) \simeq k\tau $, $\cos(k \tau) \simeq 1 $.
(See also the relation between $\tau$ and $aH$
in~(\ref{conftime}).)\label{footnote3}}  
Taking the positive
frequency solution, we can set its normalization by the
condition~(\ref{maru2}) and choose
\begin{equation}
 u_k^{(p)} = \frac{e^{-i k \tau} }{(2k)^{1/2} I}.
\end{equation}
In this case the powers of the electromagnetic fields are
\begin{equation}
 \mathcal{P}_B =  \mathcal{P}_E = \frac{k^4}{2 \pi^2 I^2 a^4},
\label{PBPEmaxwell}
\end{equation}
which redshift as $\propto a^{-4}$, 
and the spectra have a blue tilt of 
$d \ln \mathcal{P}_{B(E)} / d \ln k = 4$.

\section{Magnetogenesis During Inflation}
\label{sec:during}

Based on the discussions in the previous section, we now examine
magnetogenesis in the early universe.
We will especially focus on whether the produced magnetic field strengths reach
values inferred from gamma ray experiments that have observed
TeV gamma ray emitting blazars:
The emitted TeV-scale gamma rays interact with the extragalactic background
light and emit pairs of electrons and positrons. The produced
electron positron pairs then inverse Compton scatter off CMB photons,
producing an electromagnetic cascade.
Since magnetic fields affect the trajectories of electrons and positrons,
the (non-)detection of the secondary cascade gamma rays can
be used to measure the strength of intergalactic magnetic fields. 
Works such
as~\cite{Tavecchio:2010mk,Neronov:1900zz,Ando:2010rb,Taylor:2011bn,Takahashi:2013uoa,Finke:2013bua}
have derived lower bounds on the magnetic fields along the line of sight
towards the blazars to be of $\sim 10^{-15}$G, with an uncertainty of a
few orders. 
The correlation length of the intergalactic magnetic field
is considered to be larger than Mpc scales,
as the electron positron pairs lose energy on inverse Compton
scattering on distance scales of order Mpc.
If the magnetic field correlation length is much smaller
than a Mpc, then even stronger magnetic fields would be required to explain
the observational results. 
In the following discussions, to be specific, we aim to generate
magnetic fields with amplitude 
$ \mathcal{P}_B^{1/2} \sim 10^{-15} G $ on Mpc scales in the
present universe. 

We start by discussing magnetogenesis during inflation in this section.
As it has already been shown in previous works, we will see that
inflationary magnetogenesis is strongly constrained by the strong
coupling and backreaction problems.
We then investigate post-inflationary magnetogenesis in Section~\ref{sec:after}, 
and in Section~\ref{sec:model} a successful magnetogenesis model is presented. 
Throughout the discussions we focus on kinetic couplings~$I$ that give
rise to an effective ``horizon'' 
for a finite period of time, during which the quantum fluctuations of
the magnetic fields are enhanced. 
We also consider the coupling to scale as a power-law of the scale
factor.

\subsection{Evolution of Electromagnetic Fields}

Let us consider an inflating background with $H = H_{\mathrm{inf}} =
\mathrm{const.}$, i.e. $w = -1$.
As for the coupling, we assume that it scales as a power-law of the scale factor
until a certain time $a = a_1$, then $I$ approaches a constant value~\mbox{$I_1 $
($>0$)}, 
\begin{equation}
  I = 
 \begin{dcases}
     I_1 \left(\frac{a_1}{a}\right)^{s} 
       & \text{for $a \leq a_1$,} \\
     I_1 
       & \text{for $a > a_1$.}
 \end{dcases}
 \label{Iinf}
\end{equation}
Such a behavior can be realized by electromagnetic fields
coupled to, e.g., an inflaton or 
a spectator field, as is discussed in Section~\ref{subsec:c_during}. 
The dynamical~$I$ generates the effective horizon whose size is similar
to that of the Hubble horizon, up to the factor of~$s$.
In order for efficient magnetogenesis to happen, we suppose the power to be 
\begin{equation}
 s > \frac{1}{2},
\label{s-one-half}
\end{equation}
so that the electromagnetic mode function possesses a growing mode
outside the horizon, see~(\ref{superhlimit}).

\subsubsection{$a \leq a_1$}

The mode function~$u_k$
(Hereafter we omit the polarization index~$(p)$ as the discussions are
independent of the polarization state.)
during $a \leq a_1$ takes the
solution~$(aH_{\mathrm{inf}})^{s-\frac{1}{2}} H_{-s+ \frac{1}{2}}^{(1)}$ which
asymptotes to the positive frequency solution in the sub-horizon limit
$k \gg aH_{\mathrm{inf}} $, cf.~(\ref{subhlimit}). 
Setting the normalization from (\ref{maru2}), we obtain
\begin{equation}
 u_k = \frac{1}{2I} \left(\frac{\pi }{aH_{\mathrm{inf}}}\right) ^{1/2}
H_{-s+\frac{1}{2} }^{(1)} \left(\frac{k}{aH_{\mathrm{inf}}}\right),
\label{ukt1}
\end{equation}
up to an arbitrary phase.
When the mode is well inside the horizon, i.e. $k \gg
aH_{\mathrm{inf}}$, the mode function is approximated by
\begin{equation}
 u_k \simeq \frac{1}{(2 k)^{1/2}I}
\exp \left\{ i\left( \frac{k}{aH_{\mathrm{inf}}} + \frac{s-1}{2}
     \pi \right) \right\},
\label{ukt1sub}
\end{equation}
whose amplitude evolves as $\lvert u_k \rvert \propto a^{s}$.
In the super-horizon limit, i.e. $k \ll aH_{\mathrm{inf}}$, 
the mode function~(\ref{ukt1}) asymptotes to
\begin{equation}
 u_k \simeq   \frac{e^{i (s-\frac{1}{2} ) \pi }}{I} 
\left(\frac{\pi }{2k} \right)^{1/2}
\left\{
 \frac{1}{\Gamma (s+\frac{1}{2} )} \left(\frac{2
			   aH_{\mathrm{inf}}}{k}\right)^{-s} 
- i\frac{\Gamma(s-\frac{1}{2})}{\pi }\left( \frac{2 aH_{\mathrm{inf}}}{k}
			     \right)^{s-1 } 
\right\}.
\label{ukt1super}
\end{equation}
Together with the time dependent~$I$ in the prefactor,
the first term inside the $\{\, \}$ parentheses gives a constant mode,
while the second term gives the growing mode~$\propto a^{2 s -1}$.
We should remark that when $s$ is close to $1/2$, then
$k/a H_{\mathrm{inf}}$ needs to be extremely tiny for 
the approximation~(\ref{ukt1super}) to be valid.
In the following discussions, we suppose that $s$ is not too close to~$1/2$ 
and use (\ref{ukt1super}) for wave modes outside the horizon. 

Using the above approximations, the power spectra of the magnetic (\ref{PofB})
and electric (\ref{PofE}) fields are obtained as,
\begin{equation}
\begin{split}
  \mathcal{P}_B &\simeq
 \begin{dcases}
     \frac{k^4}{2 \pi^2 I^2 a^4} 
  \frac{\Gamma (s - \frac{1}{2} )^2}{\pi }
 \left(\frac{2 a H_{\mathrm{inf}}}{k}\right)^{2 (s-1)}
       & \text{for $k \ll a H_{\mathrm{inf}}$,} \\
     \frac{k^4}{2 \pi^2 I^2 a^4} 
       & \text{for $k \gg a H_{\mathrm{inf}}$,}
 \end{dcases}
\\
  \mathcal{P}_E &\simeq
 \begin{dcases}
     \frac{k^4}{2 \pi^2 I^2 a^4} 
\frac{\Gamma(s+\frac{1}{2})^2}{\pi }
 \left(\frac{2 a H_{\mathrm{inf}}}{k}\right)^{2 s}
       &  \mspace{28.0mu} \text{for $k \ll a H_{\mathrm{inf}}$,} \\
     \frac{k^4}{2 \pi^2 I^2 a^4} 
       &  \mspace{28.0mu} \text{for $k \gg a H_{\mathrm{inf}}$.}
 \end{dcases}
\label{PBEt1}
\end{split}
\end{equation}
From these power spectra, 
we can further compute the energy density of the electromagnetic fields. 
Contributions to the energy density from sub-horizon modes are 
renormalized, hence we take the upper limit of the integral
in~(\ref{rho_EM}) to~$a H_{\mathrm{inf}}$.
On the other hand, we write the lower limit of the integral as~$k_{\mathrm{IR}}$,
which in the case of inflationary magnetogenesis
would correspond to the wave mode that exited the horizon at the
beginning of inflation (and thus we note $k_{\mathrm{IR}} \ll a_1
H_{\mathrm{inf}}$).
We see from (\ref{PBEt1}) that the electric power spectrum is much
larger than the magnetic power on super-horizon scales.
Therefore we ignore the magnetic contributions to the energy density and
thus obtain
\begin{equation}
 \rho_{\mathrm{EM}}  \simeq
\frac{4 \Gamma(s+\frac{1}{2})^2 }{\pi^3} 
H_{\mathrm{inf}}^4
\int^{aH_{\mathrm{inf}}}_{k_{\mathrm{IR}}} \frac{dk}{k}
\left( \frac{2 a H_{\mathrm{inf}}}{k}
\right)^{2 (s - 2)},
\label{rhoEMt1}
\end{equation}
were we have extrapolated the super-horizon expression
for~$\mathcal{P}_E$ up to $k = a H_{\mathrm{inf}}$.
One can easily check that this energy density increases monotonically in time.

\subsubsection{$a > a_1$}

After the coupling~$I$ becomes time independent, i.e. $a > a_1$, 
the mode function is a sum of plane waves,
\begin{equation}
 u_k = C_+ e^{-i k (\tau - \tau_1)} + C_- e^{i k (\tau - \tau_1)}  .
\label{ukt1tend}
\end{equation}
We obtain the constants $C_+$ and $C_-$ by matching 
(\ref{ukt1tend}) and its time-derivative with the approximate
solutions (\ref{ukt1sub}), (\ref{ukt1super}) at $a = a_1$. 

For modes that are inside the horizon until $a = a_1$, i.e. 
$k \gg a_1 H_{\mathrm{inf}}$, the mode function stays as a positive frequency
solution,
\begin{equation}
 C_+ \simeq u_k (\tau_1)
\simeq \frac{1}{(2 k)^{1/2}I_1}
\exp \left\{ i\left( \frac{k}{a_1H_{\mathrm{inf}}} + \frac{s-1}{2}
     \pi \right) \right\}
, 
\qquad
C_- \simeq 0.
\label{ukt1tendsub}
\end{equation}

On the other hand for modes $k \ll a_1 H_{\mathrm{inf}}$, one finds
\begin{equation}
C_\pm \simeq  e^{i (s-\frac{1}{2})\pi}
  \frac{ \Gamma(s+\frac{1}{2})}{(8 \pi k)^{1/2} I_1}
\left( \frac{k}{2 a_1 H_{\mathrm{inf}}} \right)^{-s}
\left\{
\pm 1  + \frac{\pi }{ \Gamma (s+\frac{1}{2})^2}\left(\frac{k}{2 a_1
					     H_{\mathrm{inf}}}\right)^{2 s}
-\frac{i}{s-\frac{1}{2} } \frac{k}{2  a_1 H_{\mathrm{inf}}}
\right\}.
\label{ukt1tendsuper}
\end{equation}
Here in the $\{\, \}$ parentheses we keep the terms whose amplitudes are
much smaller than unity, as they can be important upon computing quantities
such as $\lvert u_k \rvert$, or when further connecting to mode functions in
later times. 
The time evolving~$I$
has transformed $u_k$ outside the horizon 
into a mixture of positive and negative frequency solutions.

Thus the amplitudes of the electromagnetic fields are
\begin{equation}
\begin{split}
  \mathcal{P}_B &\simeq
 \begin{dcases}
     \frac{k^4}{2 \pi^2 I_1^2 a^4} 
\frac{4 s^2 \Gamma(s - \frac{1}{2})^2}{\pi }
\left( \frac{2 a_1 H_{\mathrm{inf}}}{k} \right)^{2 (s-1)}
       & \text{for $k \ll a_1 H_{\mathrm{inf}}$,} \\
     \frac{k^4}{2 \pi^2 I_1^2 a^4} 
       & \text{for $k \gg a_1 H_{\mathrm{inf}}$,}
 \end{dcases}
\\
  \mathcal{P}_E &\simeq
 \begin{dcases}
     \frac{k^4}{2 \pi^2 I_1^2 a^4} 
   \frac{\Gamma (s + \frac{1}{2})^2}{\pi }
 \left(\frac{2 a_1 H_{\mathrm{inf}}}{k}\right)^{2s }
       &  \mspace{54.0mu} \text{for $k \ll a_1 H_{\mathrm{inf}}$,} \\ 
     \frac{k^4}{2 \pi^2 I_1^2 a^4} 
       &  \mspace{54.0mu} \text{for $k \gg a_1 H_{\mathrm{inf}}$.}
 \end{dcases}
\label{PBEt1tend}
\end{split}
\end{equation}
Here we have shown the asymptotic forms in the limit of $a \gg a_1$.
This is why the expression of $\mathcal{P}_B$ for $k \ll a_1
H_{\mathrm{inf}}$ extrapolated to $a = a_1$ differs from that in
(\ref{PBEt1}) by a factor of~$4s^2$.
Comparing the magnetic power spectra between wave modes $k \lessgtr a_1
H_{\mathrm{inf}}$, the magnetic enhancement factor from
inflationary magnetogenesis is obtained as
\begin{equation}
 \mathcal{A}_{\mathrm{inf}} = 
 \frac{4 s^2 \Gamma (s-\frac{1}{2})^2}{   \pi}
\left( \frac{2 a_1 H_{\mathrm{inf}}}{k} \right)^{2(s-1)}.
\label{Ainf}
\end{equation}
For $s > 1$, larger scales (i.e. smaller~$k$) are more significantly
enhanced as they exit the horizon earlier.
On the other hand when $1/2 < s < 1$, the growing rate of the mode
function decreases after exiting the horizon (cf. (\ref{ukt1sub}),
(\ref{ukt1super})), and thus the magnetic field strength ends up being
suppressed at large scales.

The electromagnetic energy density is obtained by integrating each of
the above power spectra over the wave modes
$k_{\mathrm{IR}} < k < a_1 H_{\mathrm{inf}}$ and $a_1 H_{\mathrm{inf}} <
k < aH_{\mathrm{inf}}$.
In the limit of $a \gg a_1$, the energy density asymptotes to
\begin{equation}
 \rho_{\mathrm{EM}} \simeq
 \rho_{\mathrm{EM}}(\tau_1) \left(\frac{a_1}{a}\right)^4 
 + \frac{H_{\mathrm{inf}}^4}{8 \pi^2},
\label{rhoEMt1tend}
\end{equation}
where $\rho_{\mathrm{EM}}(\tau_1)$ is obtained from~(\ref{rhoEMt1}).
We now have a component that decays as $\propto a^{-4}$,
plus a constant contribution of~$\mathcal{O} (H_{\mathrm{inf}}^4)$ 
from modes within the range $a_1 H_{\mathrm{inf}} < k < a H_{\mathrm{inf}}$.

\subsection{Constraints on Inflation}
\label{subsec:cons_inf}

Using the above results, we now compute the amplitude of magnetic fields
produced during inflation. 
Here we suppose that magnetogenesis happens only during inflation and
the standard Maxwell theory is recovered at $a = a_1$.
Then after magnetogenesis, the magnetic power spectrum decays
as~$a^{-4}$ until the present, cf.~(\ref{PBPEmaxwell}).  
Let us focus on a wave mode that exits the horizon before $a=a_1$, i.e. $k \ll
a_1 H_{\mathrm{inf}}$, so that the magnetic fields on this scale are
enhanced outside the horizon. 
The present amplitude of the magnetic field is 
\begin{equation}
 \mathcal{P}_B(\tau_0, k) = 4 s^2   \mathcal{P}_B(\tau_1, k)
  \left(\frac{a_1}{a_0}\right)^4,
\end{equation}
where we denote values in the present universe by the subscript~$0$,
and the 
factor~$4s^2$ arise due to the slight increase of $\abs{u_k}$
soon after $a = a_1$, see discussions below~(\ref{PBEt1tend}).

The electromagnetic energy density monotonically increases until $a =
a_1$ (cf.~(\ref{rhoEMt1})), and then it starts to decrease.
The inflation scale is restricted to satisfy $H_{\mathrm{inf}} \ll M_p$ from
observational constraints on inflationary gravitational waves, and thus the 
constant contribution of $\sim H_{\mathrm{inf}}^4$
to the energy density shown in (\ref{rhoEMt1tend}) is much smaller than the
total energy density of the inflating universe.
Therefore, the electromagnetic energy density is always smaller
than the background density if 
\begin{equation}
\rho_{\mathrm{EM}}(\tau_1) < 3 M_p H_{\mathrm{inf}}^2 
\label{energycond1}
\end{equation}
is satisfied. 

Let us now rewrite the present magnetic power spectrum in terms of the
electric spectrum at time~$\tau_1$ as
\begin{equation}
 \mathcal{P}_B(\tau_0, k) = 
\frac{s^2}{ (s-\frac{1}{2})^2}
\frac{\mathcal{P}_E (\tau_1, k) }{ H_{\mathrm{inf}}^2}
\left(\frac{k}{a_0}\right)^2
\left(\frac{a_1}{a_{\mathrm{end}}}\right)^2
\left(\frac{a_{\mathrm{end}}}{a_0}\right)^2,
\label{PB3.17}
\end{equation}
where $a_{\mathrm{end}}$ is the scale factor at the end of
inflation. 
We suppose that the universe is effectively matter-dominated after
inflation until reheating, i.e.,
\begin{equation}
 \left( \frac{H_{\mathrm{reh}}}{H_{\mathrm{inf}}} \right)^2 =
 \left( \frac{a_{\mathrm{end}}}{a_{\mathrm{reh}}} \right)^3.
\label{Hrehinf}
\end{equation}
The subscript ``reh'' denotes quantities at reheating.
The reheating scale~$H_{\mathrm{reh}}$ is related to the
entropy density~$s_{\mathrm{reh}}$ as
\begin{equation}
s_{\mathrm{reh}} = \frac{2 \pi^2}{45}g_{s*}(T_{\mathrm{reh}})
\left( \frac{90}{\pi^2}\frac{ M_p^2 H_{\mathrm{reh}}^2}{g_*
 (T_{\mathrm{reh}})} \right)^{3/4} ,
\end{equation}
and by considering the entropy to be conserved after reheating, i.e. $s
\propto a^{-3}$, 
the energy scale and redshift at reheating are related as
\begin{equation}
 \frac{H_{\mathrm{reh}}}{M_p} \approx 3 \times 10^{-63}
  \left(\frac{a_0}{a_{\mathrm{reh}}}\right)^2. 
\label{HrehMp}
\end{equation}
Here we have chosen the relativistic degrees of freedom at reheating to
be the maximum value allowed in the MSSM, $g_* = g_{s*} = 228.75$.
However we note that the numerical factor in (\ref{HrehMp}) only depends
weakly on the values of 
the relativistic degrees of freedom; e.g., $g_* = g_{s*} = 10.75$ gives
a factor~$5$ instead of~$3$ in the right hand side.
Combining (\ref{Hrehinf}) and (\ref{HrehMp}) gives
\begin{equation}
 \left(\frac{a_{\mathrm{end}}}{a_0}\right)^2
\approx 3 \times  10^{-63}
\left(\frac{H_{\mathrm{reh}}}{H_{\mathrm{inf}}}\right)^{1/3}
 \frac{M_p}{H_{\mathrm{inf}}}.
\label{rehinf63}
\end{equation}
Further using 
$1\,  G \approx 2 \times 10^{-20} \, \mathrm{GeV}^2 $ (we use the Heaviside-Lorentz
units) and $1\, \mathrm{Mpc} \approx 2 \times 10^{29}\, \mathrm{eV}^{-1}  $,
the power spectrum~(\ref{PB3.17}) can be rewritten as
\begin{equation}
 \frac{\mathcal{P}_B (\tau_0, k)}{(10^{-15}\, \mathrm{G})^2} \sim 
\frac{s^2}{ (s-\frac{1}{2})^2}
\cdot
\frac{I_1^2}{2} 
\frac{\mathcal{P}_E (\tau_1, k) }{3 M_p^2 H_{\mathrm{inf}}^2}
\cdot
\frac{1}{I_1^2}
\cdot
\left( \frac{a_1 }{a_{\mathrm{end}}} \right)^2
\left(\frac{H_{\mathrm{reh}}}{H_{\mathrm{inf}}} \right)^{1/3}
\left( \frac{k}{a_0} \, \mathrm{Mpc}  \right)^2
\frac{10^{-32} M_p}{H_{\mathrm{inf}}}.
\label{const_durINF}
\end{equation}
The factor $I_1^2 \mathcal{P}_E (\tau_1, k) /(2 \cdot 3 M_p^2 H_{\mathrm{inf}}^2)$
denotes the energy density ratio  at time~$\tau_1$ between the electric field on the
scale~$k$ and the inflating background, cf.~(\ref{rho_EM}).
This ratio
should be smaller than unity to avoid strong backreaction from the electric
field which can terminate magnetogenesis and/or inflation.
Moreover, recall from the discussions below~(\ref{maxwell}) that 
the coupling should satisfy $I_1^2
\gtrsim 1$ in order to avoid strong couplings.
Further noting that $a_1 \leq a_{\mathrm{end}}$ and $H_{\mathrm{reh}}
\leq H_{\mathrm{inf}}$, we see from~(\ref{const_durINF}) that 
an extremely low scale inflation with 
\begin{equation}
 H_{\mathrm{inf}} \ll 10^{-32} M_p \sim 10^{-5}\, \mathrm{eV}
\label{Hinfbound}
\end{equation}
is required in order to produce magnetic fields stronger 
than $10^{-15}\, \mathrm{G}$ on Mpc scales.\footnote{At first glance,
the right hand side of (\ref{const_durINF}) appears to  
blow up as $s \to 1/2$. However, as we have already mentioned, the super-horizon 
approximation~(\ref{ukt1super}) cannot be trusted when $s$ is close to $1/2$.
One can use other approximations that work well for $s = 1/2$, and see that 
in such a case the magnetic fields cannot be significantly produced,
independently of the inflation scale~$H_{\mathrm{inf}}$.}
It should also be noted that here we have discussed backreaction
from electric fields with a certain wave mode~$k$. Considering the
energy density contributions from all super-horizon modes gives an even more
stringent upper bound on~$H_{\mathrm{inf}}$.
The difficulties with inflationary magnetogenesis have 
already been pointed out in previous studies, such
as~\cite{Demozzi:2009fu,Fujita:2012rb}.\footnote{See also \cite{Ferreira:2013sqa} which
discusses magnetogenesis during significantly low scale inflation.} 
In particular, the work~\cite{Fujita:2012rb} derived a generic upper
bound on~$H_{\mathrm{inf}}$ without specifying the time evolution of the
mode function. 
Compared to their result, we arrived at a more stringent
bound~(\ref{Hinfbound}) because we have 
focused on the specific case where the coupling~$I$ scales as a
power-law of~$a$.

The strong constraint on the inflation scale can be 
understood by the fact that the magnetic fields in the present universe
traces back to possess much larger magnetic power in the earlier times,
as they redshift as $B^2 \propto a^{-4}$ after inflation. 
Thus a significant production of magnetic fields is required for
inflationary magnetogenesis that happens at high energy scales.
At the same time, higher inflation scales imply less cosmological time
for magnetogenesis,
in other words the electromagnetic vector potential
needs to be more strongly enhanced in a shorter period time for higher
inflation scales. Then the time-derivative of the vector potential would
be so large that it sources significant electric fields whose
backreaction on the inflating universe cannot be neglected.
This observation leads us to seek magnetogenesis at lower energy scales,
namely, in the post-inflationary universe.
This will be the topic of the next section.

\section{Magnetogenesis After Inflation}
\label{sec:after}

We now turn to the investigation of magnetogenesis after inflation. 
We consider the post-inflationary universe to be dominated by an
inflaton field that harmonically oscillates around its potential
minimum, behaving like pressureless matter.
The oscillating inflaton eventually decays into radiation, either
perturbatively or nonperturbatively, and reheats the universe.
During (p)reheating, the conductivity of the universe become very high,
after which the electromagnetic vector potential becomes time independent and
thus $B \propto a^{-2}$, $E=0$.
Therefore we focus on the regime between the end of inflation and
(p)reheating, during which the universe is cold and effectively
matter-dominated.

\subsection{Evolution of Electromagnetic Fields}

The background universe under consideration is matter-dominated (MD), i.e. $H
\propto a^{-3/2}$, $w=0$. As in the previous section, we study a
coupling that scales as a power-law of~$a$ during a finite period of
time,
\begin{equation}
  I = 
 \begin{dcases}
     I_1 
       & \text{for $a \leq a_2$,} \\
     I_1 \left(\frac{a_2}{a}\right)^{n } 
       & \text{for $a_2 < a \leq a_3$,} \\
     I_1 \left(\frac{a_2}{a_3}\right)^{n } \equiv I_f
       & \text{for $ a > a_3  $.}
 \end{dcases}
 \label{Imd}
\end{equation}
The coupling~$I$ asymptotes to positive constant values $I_1$ and $I_f$
in the past and future, respectively,
and the effective ``horizon'' of radius $\abs{I/\dot{I}} = \abs{n
H}^{-1}$ emerges during $a_2 < a < a_3$.
Examples of scalar field couplings with such a behavior 
are presented in~Section~\ref{subsec:c_after}.
We suppose the coupling to approach its final value~$I_f$ before reheating,
i.e., $a_3 < a_{\mathrm{reh}}$.

Under the time dependent~$I$, the mode function on super-horizon scales
is given in~(\ref{superhlimit}) with $\alpha = 2n + 1/2$.
We suppose the exponent~$n$ to
be positive, i.e.,
\begin{equation}
 n > 0,
\end{equation}
which gives rise to a growing mode $(a H)^{-4n-1} \propto a^{2 n +
\frac{1}{2} }$ for the super-horizon~$u_k$.

\subsubsection{$a \leq a_2$}

The electromagnetic theory is initially the standard Maxwell theory, 
therefore the mode function is given by the plane wave
solution as in~(\ref{ukt1tend}),
\begin{equation}
 u_k = C_+ e^{-i k (\tau - \tau_1)} + C_- e^{i k (\tau - \tau_1)}  .
\label{uktendt2}
\end{equation}
We continue our discussions from the previous section
where we studied inflationary magnetogenesis, and plug in
their results as the initial conditions for the mode functions.
The mode function for wave modes that underwent inflationary
magnetogenesis (i.e. $k \ll a_1 H_{\mathrm{inf}}$) is given
by~(\ref{ukt1tendsuper}), while 
it is simply the positive frequency solution~(\ref{ukt1tendsub}) for 
modes $k \gg a_1 H_{\mathrm{inf}}$ which 
stayed inside the horizon during inflationary magnetogenesis.

We remark that, the case where inflationary magnetogenesis did not
happen at all (or, the case where inflation itself did not happen)
can be studied by simply taking $a_1 \to 0$ 
in the following discussions, then all wave modes under consideration
would satisfy $ k \gg a_1 H_{\mathrm{inf}}$.

Let us assume the wave modes that exited the
effective/Hubble horizon during inflationary magnetogenesis to re-enter
the Hubble horizon after reheating, i.e.,
\begin{equation}
a_1 H_{\mathrm{inf}} \ll a_{\mathrm{reh}} H_{\mathrm{reh}}.
\label{assump4.4}
\end{equation}
This assumption is adopted mainly to reduce clutter in the equations;
it implies the relation $a_1 H_{\mathrm{inf}} \ll a H$ during the 
post-inflationary epoch until reheating, which will allow us to drop
many terms in the computations. 

Focusing on super-horizon modes $ k \ll a H$, 
the electromagnetic power spectra are
\begin{equation}
\begin{split}
  \mathcal{P}_B &\simeq
 \begin{dcases}
     \frac{k^4}{2 \pi^2 I_1^2 a^4} 
 \frac{4 s^2\Gamma  (s - \frac{1}{2} )^2}{\pi }
\left( \frac{2 a_1 H_{\mathrm{inf}}}{k} \right)^{2 (s-1)}
\left( 1 + \frac{2s-1}{ s}
\frac{a_1 H_{\mathrm{inf}}}{a H}  \right)^2
       & \text{for $k \ll a_1 H_{\mathrm{inf}}$,} \\
     \frac{k^4}{2 \pi^2 I_1^2 a^4} 
       & \text{for $k \gg a_1 H_{\mathrm{inf}}$,}
 \end{dcases}
\\
  \mathcal{P}_E &\simeq
 \begin{dcases}
     \frac{k^4}{2 \pi^2 I_1^2 a^4} 
 \frac{\Gamma  (s + \frac{1}{2} )^2}{\pi }
\left( \frac{2 a_1 H_{\mathrm{inf}}}{k} \right)^{2 s}
       & \mspace{228.0mu} \text{for $k \ll a_1 H_{\mathrm{inf}}$,} \\
     \frac{k^4}{2 \pi^2 I_1^2 a^4} 
       & \mspace{228.0mu} \text{for $k \gg a_1 H_{\mathrm{inf}}$,}
 \end{dcases}
\label{PBEtendt2}
\end{split}
\end{equation}
which are the same as shown in~(\ref{PBEt1tend}), except for the magnetic fields
on scales $k \ll a_1 H_{\mathrm{inf}}$.
The difference arise due to the super-horizon mode function~$u_k$ 
possessing a time evolving component
$\propto (aH)^{-1}$ which is a (slowly) growing mode in the MD universe,
cf.~(\ref{superhlimit}) and footnote~$\ref{footnote3}$. 
This component decays during inflation, and thus was neglected
in~(\ref{PBEt1tend}).
However it should be noted that this mode is always a subdominant
component under the assumption of~(\ref{assump4.4}).
We also remark that a purely positive (or negative) frequency solution
gives a constant~$\abs{u_k}$, which is why $\mathcal{P}_B$ for modes
$k\gg a_1 H_{\mathrm{inf}}$ have the same form during and after
inflation.

Integrating the electromagnetic power spectra over super-horizon 
wave modes gives the energy density 
\begin{equation}
 \rho_{\mathrm{EM}} \simeq  \rho_{\mathrm{EM}}(\tau_1)
  \left(\frac{a_1}{a}\right)^4
+ \frac{H^4}{8 \pi^2},
\label{rhoEMtendt2}
\end{equation}
which takes the same form as in (\ref{rhoEMt1tend}).
The first term on the right hand side proportional to
$\rho_{\mathrm{EM}}(\tau_1)$ denotes the electric density from wave
modes that underwent inflationary magnetogenesis, i.e., $ k_{\mathrm{IR}}
<  k < a_1 H_{\mathrm{inf}}$. 
The second term $\sim H^4$ represents the electromagnetic density mainly
sourced from waved modes entering the horizon, $k \sim aH$.

\subsubsection{$ a_2 < a \leq a_3$}

The general solution after the effective ``horizon'' emerges is shown in 
(\ref{Hankelsol}) with $\alpha = 2n + 1/2$.
By matching its super-horizon approximation~(\ref{superhlimit}) with 
(\ref{uktendt2}) at $a = a_2$, we obtain the mode function for
wave numbers $ k \ll a H$ as
\begin{multline}
 u_k \simeq
C_+ e^{-ik (\tau_2 - \tau_1)} 
\left[
1 + \frac{i}{2n+\frac{1}{2}}
\frac{k}{a_2 H_2} 
\left\{ 1 - \left(\frac{a}{a_2}\right)^{2 n+\frac{1}{2}}  \right\}
\right]
\\
+ C_- e^{ik (\tau_2 - \tau_1)} 
\left[
 1 - \frac{i}{2n+\frac{1}{2} } \frac{k}{a_2 H_2} 
\left\{  1 - \left(\frac{a}{a_2}\right)^{2 n+\frac{1}{2}}  
\right\}
\right],
\label{ukt2t3}
\end{multline}
where the constants $C_\pm$ are given in 
(\ref{ukt1tendsub}) and (\ref{ukt1tendsuper}), depending on whether
$k \gtrless a_1 H_{\mathrm{inf}}$.

When $(a/a_2)^{2n+\frac{1}{2}} \gg 1$, the electromagnetic power spectra are written as
\begin{equation}
\begin{split}
  \mathcal{P}_B &\simeq
 \begin{dcases}
     \frac{k^4}{2 \pi^2 I_1^2 a^4} 
\frac{4 s^2 \Gamma(s-\frac{1}{2})^2}{\pi }
\left( \frac{2 a_1 H_{\mathrm{inf}}}{k}\right)^{2(s-1)}
\left\{
1 + 
\frac{2 s-1}{2 s (2n + \frac{1}{2}) }
\left( \frac{a}{a_2} \right)^{2n  }
\frac{a_1 H_{\mathrm{inf}}}{a H}
\right\}^2  
 \\
\mspace{510mu} \text{for $k \ll a_1 H_{\mathrm{inf}}$,} \\
     \frac{k^4}{2 \pi^2 I_1^2 a^4} 
 \left\{
1 + \frac{1}{(2n+\frac{1}{2} )^2}
 \left(\frac{a}{a_2}\right)^{4n}
\left(\frac{k}{a H}\right)^2
\right\}
 \mspace{165mu} \text{for $k \gg a_1 H_{\mathrm{inf}}$,}
 \end{dcases}
\\
  \mathcal{P}_E &\simeq
 \begin{dcases}
     \frac{k^4}{2 \pi^2 I_1^2 a^4} \left(\frac{a}{a_2}\right)^{4n} 
\frac{\Gamma(s + \frac{1}{2})^2}{\pi }
\left(\frac{2 a_1 H_{\mathrm{inf}}}{k}\right)^{2 s}
       & \mspace{170.0mu} \text{for $k \ll a_1 H_{\mathrm{inf}}$,} \\
     \frac{k^4}{2 \pi^2 I_1^2 a^4} \left(\frac{a}{a_2}\right)^{4n} 
       & \mspace{170.0mu} \text{for $k \gg a_1 H_{\mathrm{inf}}$.}
 \end{dcases}
\label{PBEt2t3}
\end{split}
\end{equation}
Due to the time evolving~$I$, the magnetic fields now have a component
that scales as $B^2 \propto a^{4n-3}$ (given by the second terms in the
$\{\, \}$ parentheses), in addition to the normal component that decays
as $ B^2 \propto a^{-4}$. 
The former component is initially negligible, however it
can eventually dominate over the latter, depending on 
the duration of this period~$a_3/a_2$.
Hereafter, we suppose that the enhanced component $B^2
\propto a^{4n-3}$ eventually becomes dominant and assume
\begin{equation}
\left( \frac{a_3}{a_2} \right)^{2n} 
 \frac{a_1 H_{\mathrm{inf}}}{a_3 H_3} 
\gg 1.
\label{assump4.9}
\end{equation}
For modes $k \gg a_1 H_{\mathrm{inf}}$ which did not experience
inflationary magnetogenesis (or in the case where inflationary
magnetogenesis or inflation itself did not happen at all),
the assumption~(\ref{assump4.9}) is translated into
\begin{equation}
\left( \frac{a_3}{a_2} \right)^{2n} 
 \frac{k}{a_3 H_3} 
\gg 1.
\label{assump4.10}
\end{equation}

The electromagnetic energy density is enhanced
by a factor~$(a/a_2)^{2n}$ during this period. 
From the power spectra above, one finds that the energy density takes the form
\begin{equation}
 \rho_{\mathrm{EM}} \simeq
\rho_{\mathrm{EM}} (\tau_1) 
\left(\frac{a_1 }{a}\right)^4
\left( \frac{a}{a_2} \right)^{2n}
+
 \left\{ \frac{1}{2}
+ \frac{4}{3 (4n+ 1 )^2}
\right\}
\frac{H^4 }{8\pi^2}
 \left( \frac{a}{a_2} \right)^{2n},
\label{rhoEMt2t3}
\end{equation}
as $(a/a_2)^{2n}$ becomes much larger than unity.

\subsubsection{$ a >  a_3$}

In this final period before reheating, the Maxwell theory is again
recovered and the mode function reduces to a sum of plane waves,
\begin{equation}
 u_k = \widetilde{C}_+ e^{-i k (\tau - \tau_3)} + \widetilde{C}_- e^{i k
  (\tau - \tau_3)} ,
\label{ukt3treh}
\end{equation}
whose constants $\widetilde{C}_\pm$ are found by matching the solutions
(\ref{ukt2t3}) and (\ref{ukt3treh}) at $a = a_3$, giving
\begin{multline}
  \widetilde{C}_\pm \simeq
\frac{C_+}{2} e^{- ik ( \tau_2 - \tau_1 )} 
\left[
1 \pm \left(\frac{a_3}{a_2}\right)^{2n}
+ \frac{i}{2n+\frac{1}{2}  }
\frac{k}{a_2 H_2}
\left\{
1
- \left(\frac{a_3}{a_2}\right)^{2n+\frac{1}{2} }
\right\}
\right]
\\
+ \frac{C_-}{2}  e^{ik (\tau_2 - \tau_1)}
\left[
1 \mp  \left(\frac{a_3}{a_2}\right)^{2n} 
-\frac{i}{2n+\frac{1}{2}  }\frac{k}{a_2 H_2}
\left\{
1 - \left(\frac{a_3}{a_2}\right)^{2n + \frac{1}{2}} 
\right\}
\right].
\end{multline}

When $ a \gg a_3 $, the power spectra for the super-horizon modes $ k
\ll a H$ are
\begin{equation}
\begin{split}
  \mathcal{P}_B &\simeq
 \begin{dcases}
\frac{k^4}{2 \pi^2 I_f^2 a^4}
\frac{16  \Gamma (s+\frac{1}{2})^2}{\pi }
\left(\frac{2 a_1 H_{\mathrm{inf}}}{k}\right)^{2(s-1)}
\left( \frac{a_3}{a_2} \right)^{2n}
\left(\frac{a_1 H_{\mathrm{inf}}}{a H}\right)^2
        \\
\mspace{480mu} \text{for $k \ll a_1 H_{\mathrm{inf}}$,} \\
     \frac{k^4}{2 \pi^2 I_f^2 a^4} 
\left(\frac{a_3}{a_2}\right)^{2n} 
\left( \frac{2k}{a H} \right)^2
\mspace{276mu}  \text{for $k \gg a_1 H_{\mathrm{inf}}$,}
 \end{dcases}
\\
  \mathcal{P}_E &\simeq
 \begin{dcases}
     \frac{k^4}{2 \pi^2 I_f^2 a^4} 
\left( \frac{a_3}{a_2} \right)^{2 n}
\frac{ \Gamma(s + \frac{1}{2} )^2}{\pi }
\left( \frac{2 a_1 H_{\mathrm{inf}}}{k} \right)^{2 s}
       & \mspace{141.0mu} \text{for $k \ll a_1 H_{\mathrm{inf}}$,} \\
     \frac{k^4}{2 \pi^2 I_f^2 a^4} 
\left(\frac{a_3}{a_2}\right)^{2 n}
       & \mspace{141.0mu} \text{for $k \gg a_1 H_{\mathrm{inf}}$.}
 \end{dcases}
\label{PBEt3treh}
\end{split}
\end{equation}
Note that here we have used the final value of the coupling~$I_f$, cf.~(\ref{Imd}).
Even for modes with $k \gg a_1 H_{\mathrm{inf}}$
which did not undergo magnetogenesis during inflation,
the mode function~$u_k$ has been transformed into a mixture of positive
and frequency solutions during the post-inflationary evolution of~$I$,
and now $u_k$ is dominated by the growing mode $\propto a^{1/2}$.

Comparing $\mathcal{P}_B$ for modes $k \gg a_1 H_{\mathrm{inf}}$ with
the power spectrum in the standard 
Maxwell theory (\ref{PBPEmaxwell}) at reheating,
the magnetic enhancement factor from post-inflationary magnetogenesis reads
\begin{equation}
 \mathcal{A}_{\mathrm{post}} = 
\left(\frac{a_3}{a_2}\right)^{2n}
\left( \frac{2 k}{a_{\mathrm{reh}} H_{\mathrm{reh}}} \right)^2.
\label{Apost}
\end{equation}
We note that this enhancement is for wave modes that stay outside the
Hubble (and effective) horizon throughout the post-inflationary
magnetogenesis.
In contrast to inflationary magnetogenesis~(\ref{Ainf}), here the
enhancement is stronger at smaller scales (larger~$k$).

Magnetic fields on wave modes $k \ll a_1 H_{\mathrm{inf}}$ are enhanced 
both during and after inflation. The net enhancement factor from 
magnetogenesis in the two regimes is
\begin{equation}
\begin{split}
 \mathcal{A}_{\mathrm{full}} &=
\frac{16   \Gamma (s+\frac{1}{2} )^2}{\pi }
\left( \frac{2 a_1 H_{\mathrm{inf}}}{k} \right)^{2(s-1)}
\left(\frac{a_3}{a_2}\right)^{2n}
\left( \frac{a_1 H_{\mathrm{inf}}}{a_{\mathrm{reh}} H_{\mathrm{reh}}}
\right)^2
\\ &=
 \mathcal{A}_{\mathrm{inf}} \mathcal{A}_{\mathrm{post}}
\left( \frac{2 s- 1}{2s} \right)^2
\left(\frac{a_1 H_{\mathrm{inf}}}{k}\right)^2.
\label{Afull}
\end{split}
\end{equation}
Interestingly the combination of inflationary and
post-inflationary magnetogenesis 
gives rise to a much stronger enhancement than the simple product of the
individual magnetogenesis scenarios.

Finally, the electromagnetic energy density when $a \gg a_3$ is
\begin{equation}
 \rho_{\mathrm{EM}} \simeq
 \rho_{\mathrm{EM}}(\tau_1) \left(\frac{a_1}{a}\right)^4
\left( \frac{a_3}{a_2} \right)^{2n}
+ \frac{11}{6} \frac{H^4}{8 \pi^2} \left(\frac{a_3}{a_2} \right)^{2n}.
\label{rhoEMt3treh}
\end{equation}

\subsection{Constraints on Reheating}
\label{subsec:con_reh}

In this subsection we analyze the magnetic fields produced from
post-inflationary magnetogenesis alone. We will see that
post-inflationary magnetogenesis is also strictly constrained by the
strong coupling and backreaction problems,
and that a significant production of large-scale magnetogenesis would require a
reheating temperature that is extremely low.\footnote{Here we
discuss reheating constraints on post-inflationary magnetogenesis,
however we note that inflationary magnetogenesis can also impose
constraints on reheating. See \cite{Ringeval:2013hfa,Demozzi:2012wh}.}

As is explained above (\ref{PBEtendt2}), we can focus solely on
post-inflationary magnetogenesis by taking $a_1 \to 0$ and studying 
the wave modes $ k \gg a_1 H_{\mathrm{inf}}$ in the above discussions.
Hence for all wave modes we only need to consider the set of constants
$\abs{C_+ }^2 = 1/2 k I_1^2$, $C_- = 0$, as shown in~(\ref{ukt1tendsub}).

Correspondingly, the electromagnetic energy densities (\ref{rhoEMtendt2}),
(\ref{rhoEMt2t3}), and (\ref{rhoEMt3treh})\footnote{The magnetic power
spectrum for small wave numbers that do not satisfy the 
condition~(\ref{assump4.10}) can be different from $\mathcal{P}_B$
for $k \gg a_1 H_{\mathrm{inf}}$ shown in~(\ref{PBEt3treh}). However
the contribution to the energy density from such small wave numbers is
negligibly tiny, and so
the result from (\ref{rhoEMt3treh}), $\rho_{\mathrm{EM}} \simeq (11 H^4/48
\pi^2) (a_3/a_2)^{2n}$, is unaffected.} 
only have the components proportional to~$H^4$. 
In particular, the energy density ratio between the electromagnetic
fields and the background universe scales as
\begin{equation}
\frac{\rho_{\mathrm{EM}} }{3 M_p^2 H^2}\simeq
\begin{dcases}
 \frac{H^2}{24 \pi^2  M_p^2} 
\propto a^{-3}
& \text{for $a \leq a_2$,} \\
 \left( \frac{1}{2}
+ \frac{4}{3 (4n+ 1 )^2}
\right)
\frac{H^2 }{24\pi^2  M_p^2 }
 \left( \frac{a}{a_2} \right)^{2n}
\propto a^{2n-3}
& \text{for $a_2 < a \leq a_3$,} \\
 \frac{11}{6} \frac{H^2}{24 \pi^2 M_p^2 } \left(\frac{a_3}{a_2}
 \right)^{2n}
\propto a^{-3}
& \text{for $a_3 < a  $.}
\end{dcases}
\label{energyratio-post}
\end{equation}
Considering a sub-Planckian universe, i.e. $H \ll M_p$, the
electromagnetic backreaction can become significant only if $n > 3/2$
so that the ratio increases during $a_2 < a \leq a_3$. In other words,
the backreaction is tiny throughout magnetogenesis if
\begin{equation}
 \frac{\rho_{\mathrm{EM}}(\tau_3)}{3 M_p^2 H_3^2} \ll 1
\label{cond4.18}
\end{equation}
is satisfied.

We focus on scales that re-enter the Hubble horizon after reheating,
i.e. $k \ll a_{\mathrm{reh}} H_{\mathrm{reh}}$, 
and also satisfy the condition~(\ref{assump4.10})
so that the magnetic fields are significantly enhanced on this scale.
Then considering that the magnetic power after reheating scales as
$\mathcal{P}_B \propto a^{-4}$, the present value reads
\begin{equation}
 \mathcal{P}_B (\tau_0, k) = 
     \frac{k^4}{2 \pi^2 I_f^2 a_{\mathrm{reh}}^4} 
\left(\frac{a_3}{a_2}\right)^{2n} 
\left( \frac{2k}{a_{\mathrm{reh}} H_{\mathrm{reh}}} \right)^2
\left( \frac{a_{\mathrm{reh}}}{a_0} \right)^4.
\end{equation}
Expressing $a_{\mathrm{reh}}$ in terms of $H_{\mathrm{reh}}$
using~(\ref{HrehMp}), and also from the second line
of~(\ref{energyratio-post}), we find 
\begin{equation}
 \frac{\mathcal{P}_B (\tau_0, k)}{(10^{-15}\, \mathrm{G})^2} \sim
 \frac{3 (4n+1 )^2}{3 (4n+1)^2 + 8}
\cdot
 \frac{\rho_{\mathrm{EM}}(\tau_3)}{3 M_p^2 H_3^2}
\cdot
\frac{1}{I_f^2}
\cdot
\left( \frac{H_{\mathrm{reh}}}{H_3} \right)^2
\left( \frac{k}{a_0} \, \mathrm{Mpc} \right)^6
\left( \frac{ 10^{-23}\, \mathrm{MeV}}{H_{\mathrm{reh}}} \right)^3.
\label{const_postINF}
\end{equation}
Here, note that the condition~(\ref{cond4.18}) is required to avoid
significant backreaction, $I_f^2 \gtrsim 1$ to avoid strong couplings in
the theory, and $H_\mathrm{reh} \leq H_3$.
Thus we see that requiring the generated magnetic fields 
to be stronger than $\sim  10^{-15}\,
\mathrm{G}$ on Mpc scales bounds the reheating scale from above as
\begin{equation}
 H_{\mathrm{reh}} \ll 10^{-23}\, \mathrm{MeV}.
\label{Hrehbound}
\end{equation}
Such a low reheating scale is inconsistent with Big Bang
Nucleosynthesis (BBN) which requires the reheating temperature to be
at least about $ 5\,
\mathrm{MeV}$~\cite{Kawasaki:1999na,Kawasaki:2000en,Hannestad:2004px},  
i.e., 
\begin{equation}
 H_{\mathrm{reh}} \gtrsim 10^{-20}\, \mathrm{MeV}.
\label{BBNconst}
\end{equation}

Thus we see that post-inflationary magnetogenesis alone cannot produce
large magnetic fields on cosmological scales without having significant
backreaction on the expanding universe, or going into the strong
coupling regime. 
The magnetic enhancement is stronger for lower reheating scales, as 
is seen from the enhancement
factor~$\mathcal{A}_{\mathrm{post}}$ in~(\ref{Apost}). 
Furthermore, the enhanced electromagnetic power spectra have larger amplitudes at
smaller scales (larger~$k$); the spectral indices are  $d
\ln \mathcal{P}_B / d \ln k = 6$, $d \ln \mathcal{P}_E / d \ln k = 4$,
cf.~(\ref{PBEt3treh}).  
(This is in contrast to inflationary magnetogenesis which can produce 
red-tilted spectra.)
As a consequence, the dominant contribution to the electromagnetic energy
density comes from wave modes entering the horizon, i.e.~$k^{-1} = (a H)^{-1}$.
Since the comoving Hubble radius $(aH)^{-1}$ grows in an MD universe,
post-inflationary  magnetogenesis is pushed to later times
in order to avoid the backreaction issue, and we have seen that
this conflicts with BBN.

Before ending this section we should remark that,
even though we have been trying to avoid electromagnetic backreaction,
their actual effects on the MD universe is unclear. 
The work~\cite{Kanno:2009ei} studied a model of inflationary
magnetogenesis and showed that electromagnetic backreaction
significantly suppresses the production of magnetic fields during inflation.
It would be interesting to examine whether backreaction is actually
disastrous for post-inflation magnetogenesis as well.

\section{Large Magnetic Fields from Two Step Magnetogenesis}
\label{sec:model}

We have seen in the previous sections the difficulties of the individual inflationary
and post-inflationary magnetogenesis.
Inflationary magnetogenesis imposed a severe upper bound on the
inflation scale, while post-inflationary magnetogenesis could not be
accomplished without spoiling BBN.

In this section we show that efficient production of large-scale magnetic fields 
can be achieved by combining the inflationary and post-inflationary
magnetogenesis. 
This is possible due to the magnetogenesis in the two epochs working in
different ways, for instance in terms of the scale dependencies of the
produced electromagnetic spectra.
One advantage of the two step magnetogenesis is that,
the post-inflationary magnetic enhancement relaxes the generation
of electric as well as magnetic fields during inflation.
Another advantage is that, since inflationary magnetogenesis is
typically more effective at larger scales 
(cf.~(\ref{Ainf})), it can compensate for the 
small scale (though super-horizon) electromagnetic fields intensively produced from
post-inflationary magnetogenesis. 
Therefore the inflationary and post-inflationary magnetogenesis
complement each other to evade the backreaction problem.
Moreover, as we have seen in (\ref{Afull}), 
post-inflation magnetogenesis works even more effectively for wave modes
that has already underwent inflationary magnetogenesis. 

\subsection{Magnetic Fields and Energy Bounds}
\label{subsec:energy_bounds}

The basic formulae are already laid out in the previous sections, 
however a few things are worth noting.
For wave modes that experience magnetic enhancement both during and
after inflation, i.e. $k \ll a_1 H_{\mathrm{inf}}$, 
the magnetic power in the present universe 
is obtained by using (\ref{PBEt3treh}) as
\begin{equation}
\begin{split}
 \mathcal{P}_B(\tau_0, k) 
& =  \mathcal{P}_B(\tau_{\mathrm{reh}}, k)
  \left(\frac{a_{\mathrm{reh}}}{a_0}\right)^4
\\
& = 
\frac{k^4}{2 \pi^2 I_f^2 a_{\mathrm{reh}}^4}
\frac{16\Gamma(s+\frac{1}{2})^2}{\pi }
\left(\frac{2 a_1 H_{\mathrm{inf}}}{k}\right)^{2(s-1)}
\left(\frac{a_3}{a_2}\right)^{2n}
\left(\frac{a_1 H_{\mathrm{inf}}}{a_{\mathrm{reh}}
 H_{\mathrm{reh}}}\right)^2
  \left(\frac{a_{\mathrm{reh}}}{a_0}\right)^4.
\label{PBnow}
\end{split}
\end{equation}
We can further rewrite the expression using (\ref{HrehMp}) as
\begin{equation}
 \frac{\mathcal{P}_B(\tau_0, k) }{(10^{-15}\, \mathrm{G})^2}
\sim
10^{-94}
\left( \frac{2 a_1 H_{\mathrm{inf}}}{k} \right)^{2s}
\left(\frac{a_3}{a_2}\right)^{2n}
 \frac{\Gamma(s+\frac{1}{2} )^2   }{ I_f^2}
\left(\frac{k}{a_0}\, \mathrm{Mpc}\right)^6
\frac{10^{-20}\, \mathrm{MeV}}{H_{\mathrm{reh}}}.
\label{5.2rewrite}
\end{equation}
Considering $I_f^2 \gtrsim 1$ and the BBN constraint~(\ref{BBNconst}),
one sees that $B \sim 10^{-15}\, \mathrm{G}$ on Mpc scales can be
produced if the combined inflationary/post-inflationary effect gives
$( 2 a_1 H_{\mathrm{inf}}/k )^{2s} (a_3/a_2 )^{2n} \gtrsim 10^{94}$.

As we have seen in the previous sections, the electromagnetic energy
density basically consists of two components which are respectively proportional to
$\rho_{\mathrm{EM}}(\tau_1)$ and $H^4$, each multiplied by some powers of
the scale factor. 
(We already discussed the $H^4$ component in Section~\ref{subsec:con_reh}, 
and the electromagnetic density during inflation in Section~\ref{subsec:cons_inf}.)
The energy density ratio between the electromagnetic fields and the background
$\rho_{\mathrm{EM}} / 3 M_p^2 H^2$ grows
monotonically during $a < a_1$, while it decreases during the periods
$a_1 < a < a_2$ and $a_3 < a < a_{\mathrm{reh}}$.
During $ a_2 < a < a_3$, the two components of~$\rho_{\mathrm{EM}}$ may
either grow or decay, depending on the value of the exponent~$n$.
Therefore we find that the electromagnetic backreaction is always tiny
if the energy ratio is small at times $\tau_1 $ and $\tau_3$, i.e., 
\begin{align}
 \frac{\rho_{\mathrm{EM}}(\tau_1)}{3 M_p^2 H_{\mathrm{inf}}^2} &\ll 1, 
\label{ratioat1}
\\
 \frac{\rho_{\mathrm{EM}}(\tau_3)}{3 M_p^2 H_3^2}  &\ll 1.
\label{ratioat3}
\end{align}
The energy condition at $\tau_1$~(\ref{ratioat1}) constrains the magnetic field
amplitude in a similar fashion as in~(\ref{const_durINF}), except for
that now we have an extra enhancement for the magnetic amplitude of
\begin{equation}
\frac{ \mathcal{A}_{\mathrm{full}}}{ \mathcal{A}_{\mathrm{inf}}}
= 
\mathcal{A}_{\mathrm{post}}
\left( \frac{2 s- 1}{2s} \right)^2
\left(\frac{a_1 H_{\mathrm{inf}}}{k}\right)^2
= \left( \frac{2s-1}{s} \right)^2 \left(\frac{a_3}{a_2}\right)^{2n}
\left( \frac{a_1 H_{\mathrm{inf}}}{a_{\mathrm{reh}} H_{\mathrm{reh}}} \right)^2,
\end{equation}
which arise from the post-inflationary magnetogenesis. This extra term
can significantly weaken the
bound~(\ref{Hinfbound}) on the inflation scale. 

The bound at~$\tau_3$~(\ref{ratioat3}) consists of two conditions, namely,
\begin{gather}
  \left(
\frac{1}{2} + \frac{4}{3 (4n+ 1 )^2}
\right)
\frac{H_3^2 }{24 \pi^2  M_p^2 }
 \left( \frac{a_3}{a_2} \right)^{2n}
\ll 1,
\label{eq5.4}
\\ 
\frac{\rho_{\mathrm{EM}} (\tau_1) }{3 M_p^2 H_3^2}
\left(\frac{a_1 }{a_3}\right)^4
\left( \frac{a_3}{a_2} \right)^{2n}
\ll 1.
\label{eq5.5}
\end{gather}
The former condition~(\ref{eq5.4}) restricts the magnetic fields as
in~(\ref{const_postINF}), but now with an extra factor of 
\begin{equation}
\frac{ \mathcal{A}_{\mathrm{full}}}{ \mathcal{A}_{\mathrm{post}}}
= 
 \mathcal{A}_{\mathrm{inf}}
\left( \frac{2 s- 1}{2s} \right)^2
\left(\frac{a_1 H_{\mathrm{inf}}}{k}\right)^2
=
\frac{\Gamma(s+\frac{1}{2})^2}{\pi }
\left( \frac{2 a_1 H_{\mathrm{inf}}}{k} \right)^{2s}
\end{equation}
in the right hand side.
The extra enhancement due to the inflationary magnetogenesis 
weakens the constraint~(\ref{Hrehbound}) on reheating.

The latter bound~(\ref{eq5.5}) is a constraint that arise only in the
combined case of inflationary and post-inflationary magnetogenesis;
it represents the constraint on large-scale electric fields that were
produced during inflation, and further enhanced by the post-inflationary
magnetogenesis. 
Focusing on this bound, the magnetic power~(\ref{PBnow}) can be
rewritten as
(recall that $\rho_{\mathrm{EM}}(\tau_1)$ can be obtained by taking
$a=a_1$ in~(\ref{rhoEMt1})),\footnote{In obtaining the right hand side of~(\ref{PB0_twostep}),
we have assumed $s \neq 2$. 
The case with $s = 2$ is obtained by simply replacing the terms
in the first line by
\begin{equation}
2(s-2)
\left\{
1- \left( \frac{k_{\mathrm{IR}}}{a_1 H_{\mathrm{inf}}}\right)^{2 (s-2)}  
\right\}^{-1}
\, \longrightarrow \, 
\left\{  \ln  \left(\frac{a_1 H_{\mathrm{inf}}}{k_{\mathrm{IR}}} \right)\right\}^{-1}.
\end{equation}
}
\begin{multline}
 \frac{ \mathcal{P}_B(\tau_0, k) }{(10^{-15}\, \mathrm{G})^2} \sim
2(s-2)
\left\{
1- \left( \frac{k_{\mathrm{IR}}}{a_1 H_{\mathrm{inf}}}\right)^{2 (s-2)}  
\right\}^{-1}
\left( \frac{k_{\mathrm{IR}}}{k} \right)^{2(s-2)}
\\
\cdot
 \frac{\rho_{\mathrm{EM}} (\tau_1) }{3 M_p^2 H_3^2}
\left(\frac{a_1 }{a_3}\right)^4
\left( \frac{a_3}{a_2} \right)^{2n}
\cdot
\frac{1}{I_f^2}
\cdot
\frac{a_3 }{a_{\mathrm{reh}}} 
\left( \frac{k}{a_0}\, \mathrm{Mpc} \right)^2
\frac{10^{-10}\, \mathrm{MeV}}{H_{\mathrm{reh}}}.
\label{PB0_twostep}
\end{multline}
Since we are interested in wave modes lying in the range $k_{\mathrm{IR}} \ll k
\ll a_1 H_{\mathrm{inf}}$, the first line of the right hand
side is smaller than unity.
The right hand side is also suppressed by 
the energy condition~(\ref{eq5.5}), the requirement~$I_f^2
\gtrsim 1$ for avoiding strong coupling, and $a_3 < a_{\mathrm{reh}}$.
In order to produce $B \sim 10^{-15}\, \mathrm{G}$
on Mpc scales, the reheating scale~$H_{\mathrm{reh}}$ should be smaller
than $\sim 10^{-10} \, \mathrm{MeV}$, 
or in terms of the reheating temperature less than a few hundred GeV.
Thus the reheating scale cannot be arbitrary high, however we stress
that the bound is now significantly weakened compared
to the case of post-inflation magnetogenesis alone~(\ref{Hrehbound}),
and that the two step magnetogenesis can be completed before BBN.

\subsection{Case Study}

Let us now demonstrate that the two step magnetogenesis can actually 
produce large magnetic fields, even with high scale inflation.
We examine the evolution of the electromagnetic spectra 
under a fixed set of parameters that allow efficient magnetogenesis. 

The example parameters are chosen as follows:
We set the inflation scale to 
$H_{\mathrm{inf}} = 10^{-6} M_p \sim 10^{12}\, \mathrm{GeV}$, 
and the reheating scale at $H_{\mathrm{reh}} = 10^{-18}\, \mathrm{MeV}$
(which corresponds to
$T_{\mathrm{reh}} \approx 50\, \mathrm{MeV}$
in terms of temperature).
The infrared cutoff of the energy density integral~(\ref{rhoEMt1}),
corresponding to the wave mode that exits the horizon at the beginning
of inflation, is set to
$k_{\mathrm{IR}}/a_0 = 10^{-2} H_0 \sim 10^{-6} \, \mathrm{Mpc}^{-1}$,
i.e., during inflation the wave mode~$k_{\mathrm{IR}}$ exited the
horizon about $5$~e-foldings before the mode corresponding to 
the present Hubble radius did.

The coupling~$I$ scales with 
$s=5/2$ during inflation as in~(\ref{Iinf}), 
$ n = 6 $ after inflation as~(\ref{Imd}),
then settles down to its final value~$I_f = 1$.
We set the scale factor~$a_1$ by 
$a_1 H_{\mathrm{inf}} / a_0 = 10^{4}\, \mathrm{Mpc}^{-1}$,
which by using~(\ref{rehinf63}) can be rephrased as
the time evolution of~$I$ to first terminate
at roughly $30$~e-foldings before the end of inflation.
The second phase of $I$ evolution in the post-inflation universe
lasts for a period of $a_3 / a_2 \approx 1.4 \times  10^6$,
where $a_3$ is related to the scale factor at reheating by
$ a_{\mathrm{reh}}/a_3 = 10 $.
In terms of the energy scale, these corresponds to
$H_2 \sim 10^{-7}\, \mathrm{MeV}$ and $H_3 \sim 10^{-17} \, \mathrm{MeV}$. 

This set of parameters gives magnetic fields of 
$\mathcal{P}_B (\tau_0, k)^{1/2} \sim 10^{-15}\, \mathrm{G}$
on scales $k / a_0 \sim 1\, \mathrm{Mpc}^{-1}$
(which is easily seen from the formula~(\ref{5.2rewrite})), while
satisfying the energy bounds (\ref{ratioat1}) and (\ref{ratioat3}) as 
$  \rho_{\mathrm{EM}}(\tau_1)/3 M_p^2 H_{\mathrm{inf}}^2 \sim 10^{-3}$, 
$ \rho_{\mathrm{EM}}(\tau_3)/3 M_p^2 H_3^2 \sim 10^{-2}$.
\\

\begin{figure}[t]
\centering
\subfigure[$a = 10^{-2} a_1 $]{%
  \includegraphics[width=.47\linewidth]{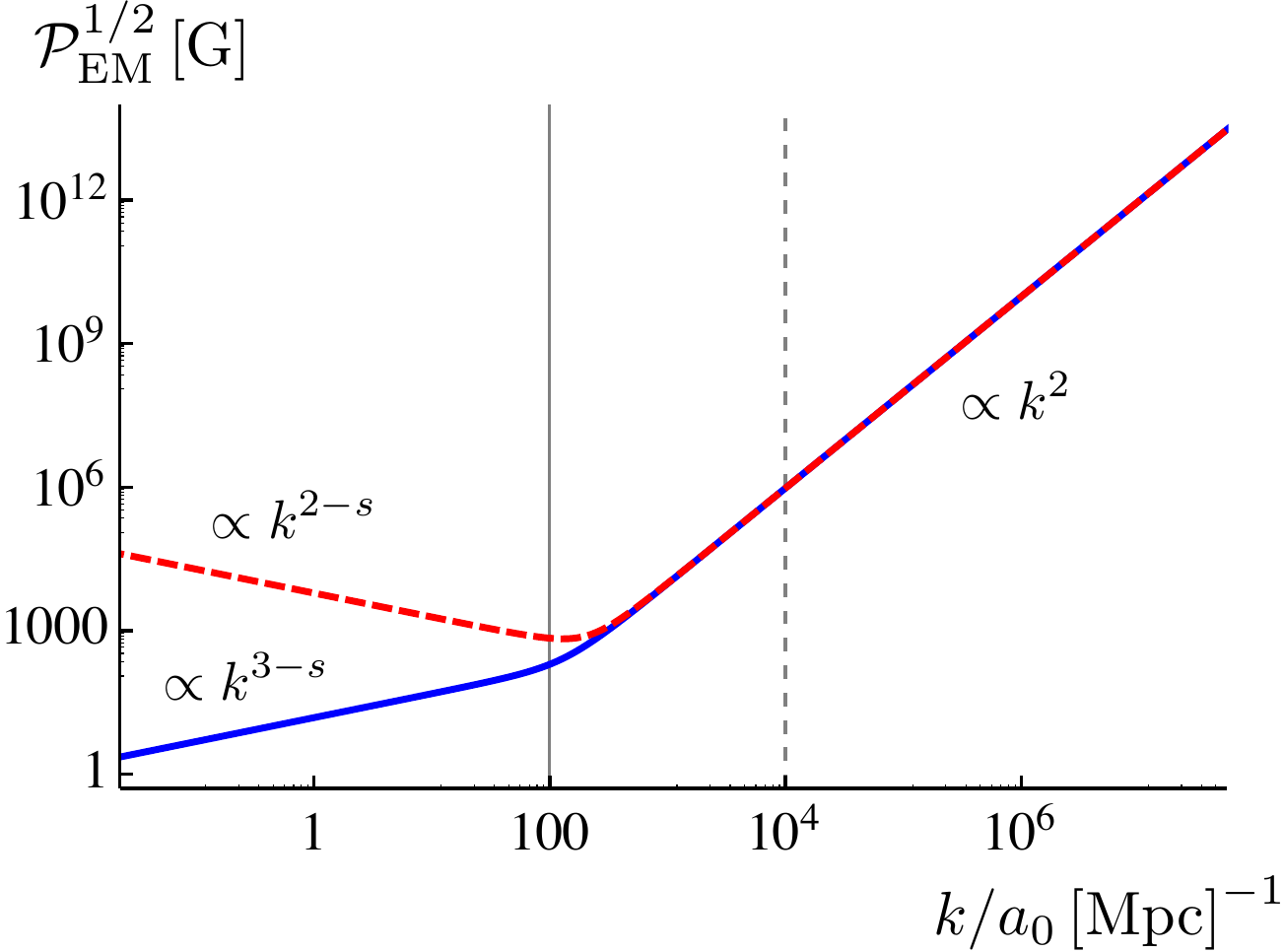}
  \label{fig:sp_01}}
\quad
\subfigure[$a \approx 3 \times 10^{8} a_1 $ ($< a_{\mathrm{end}}$)]{%
  \includegraphics[width=.47\linewidth]{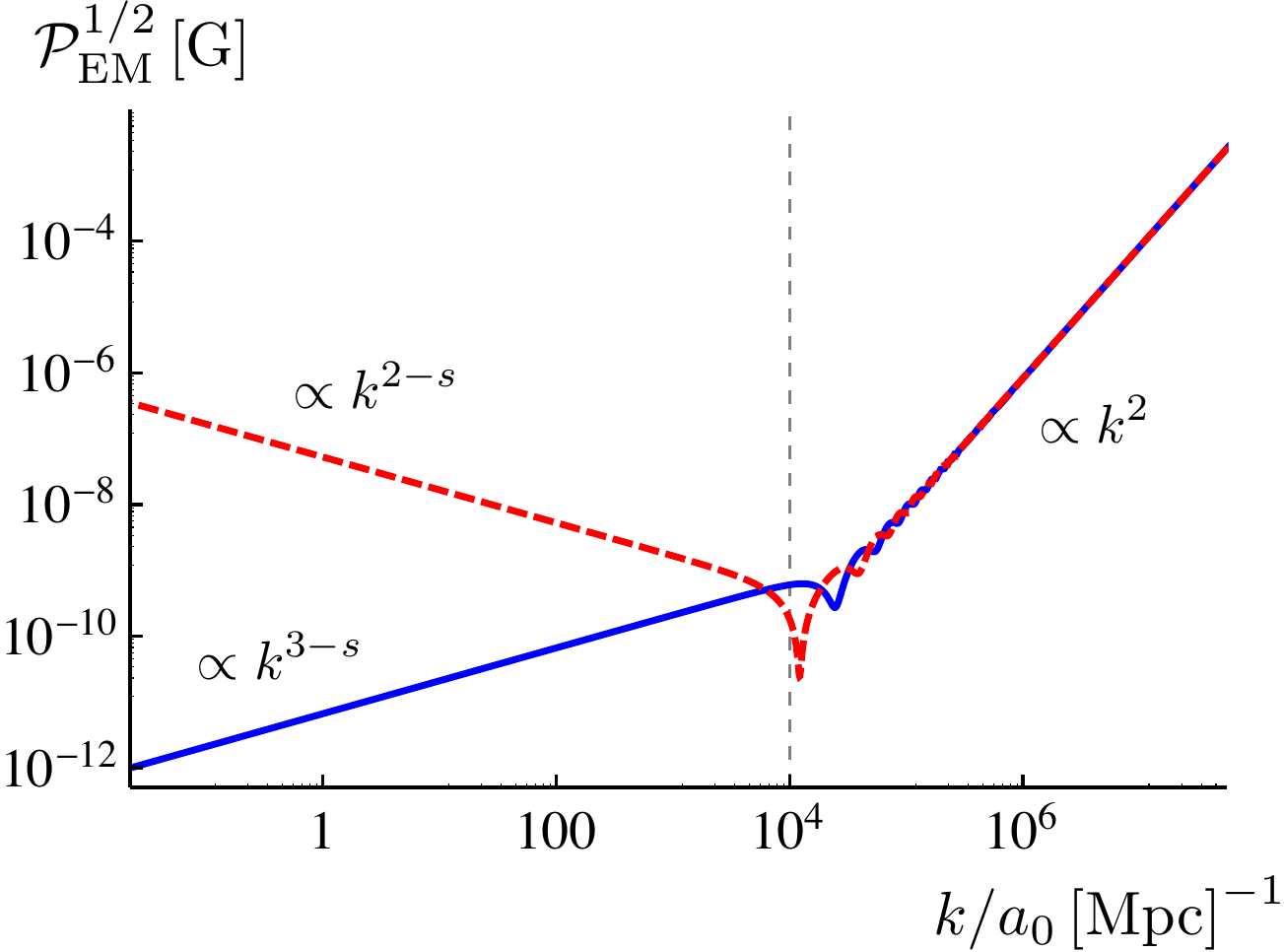}
  \label{fig:sp_12}}
\subfigure[$a \approx 2 a_2 $ ($< a_3$)]{%
  \includegraphics[width=.47\linewidth]{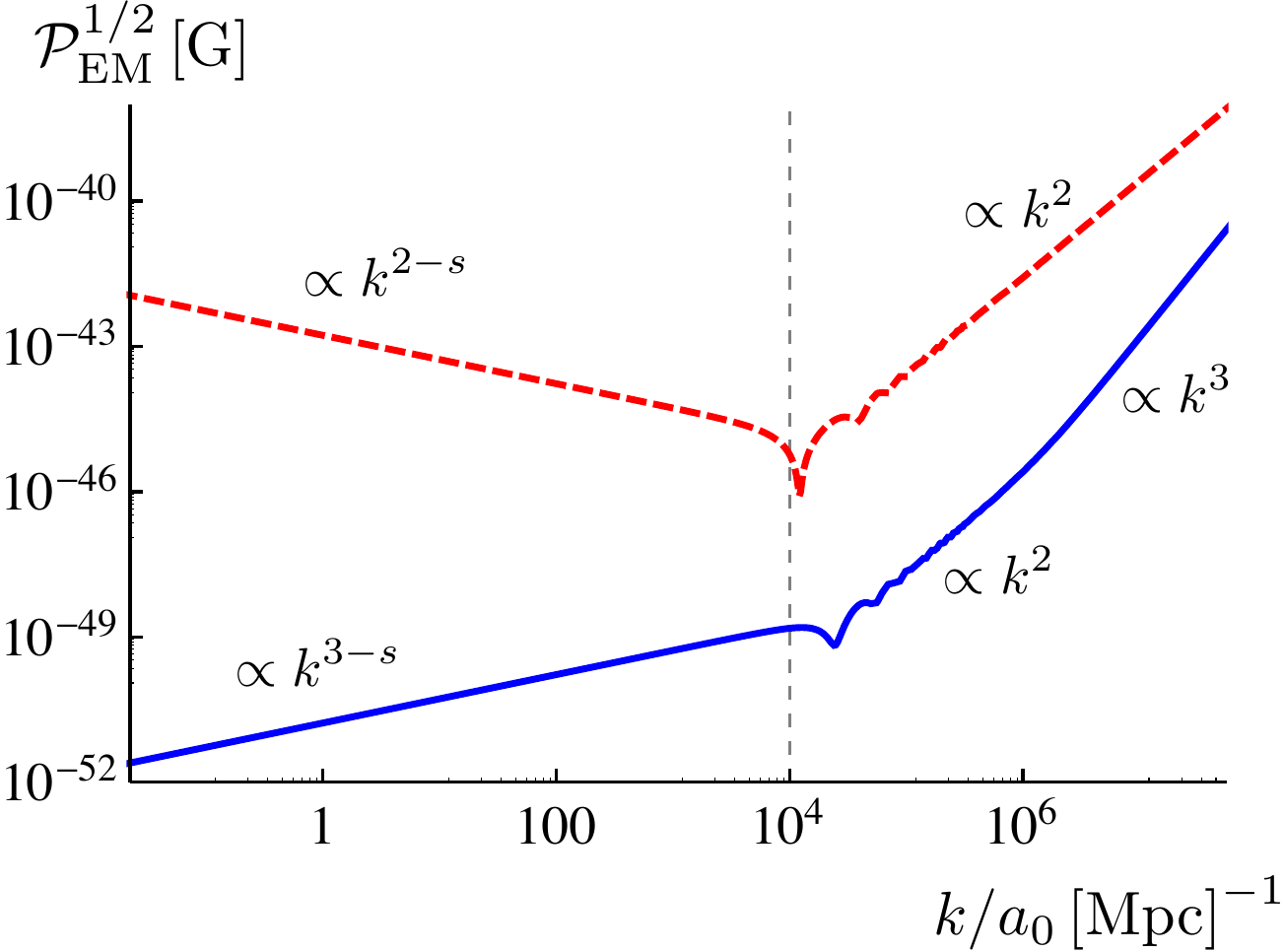}
  \label{fig:sp_23}}
\quad
\subfigure[$a = a_{\mathrm{reh}}$]{%
  \includegraphics[width=.47\linewidth]{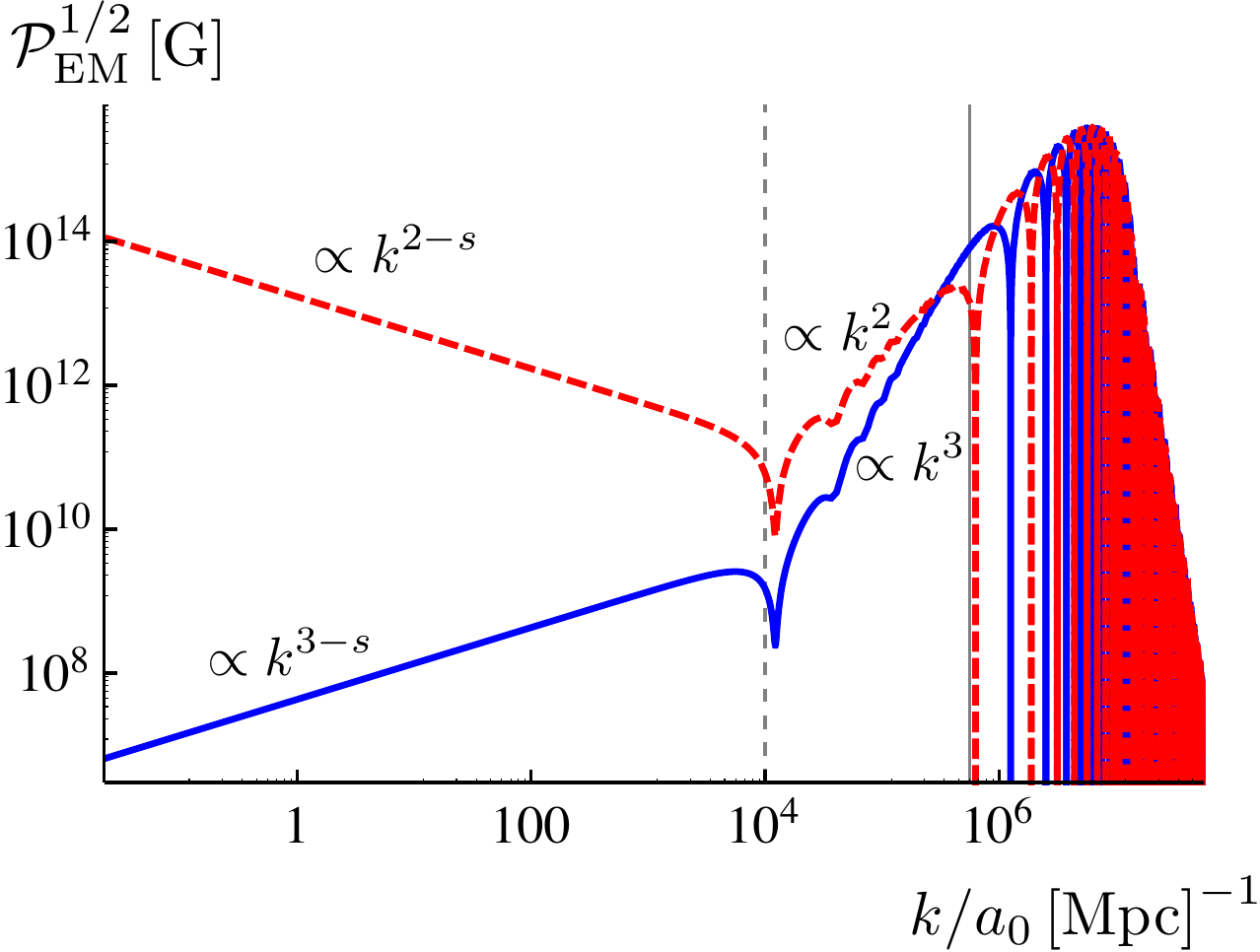}
  \label{fig:sp_reh}}
\caption{Electromagnetic power spectra at different
 times. $\mathcal{P}_B^{1/2}$ is shown as the blue solid lines, and
 $\mathcal{P}_E^{1/2}$ by the red dashed lines. The solid vertical lines
 denote the Hubble 
 radius at each time (i.e. $k = aH$), while the dashed vertical lines
 represent the wave mode~$k =  a_1 H_{\mathrm{inf}}$.} 
\label{fig:EMspectra}
\end{figure}

In Figure~\ref{fig:EMspectra} we plot the electromagnetic power spectra
at different times.
The blue solid lines denote the magnetic
power spectra~$\mathcal{P}_B^{1/2}$, and the red dashed lines are the electric
power~$\mathcal{P}_E^{1/2}$. 
Upon plotting the spectra, 
we have set the mode function~$u_k$ as the positive frequency
solution~(\ref{ukt1}) during $a < a_1$, 
and then used the exact solution for~$u_k$ (\ref{Hankelsol}) at each epoch.  
The results of course match with the approximate expressions presented
in the previous sections, except for at intermediate scales such as $k
\sim a_1 H_{\mathrm{inf}}$, where we switch between different
limits used in the approximations.
The dashed vertical lines in the figures represent the wave mode $k = a_1
H_{\mathrm{inf}}$, while the solid vertical lines in
Figures~$\ref{fig:sp_01}$ and $\ref{fig:sp_reh}$ show the Hubble
radius, i.e. $k = aH$. In Figures~$\ref{fig:sp_12}$ and
$\ref{fig:sp_23}$, the displayed wave modes are all outside the horizon.

Figure~\ref{fig:sp_01} shows the electromagnetic power spectra 
at $a = 10^{-2} a_1 $. 
Inflationary magnetogenesis lifts the spectra on super-horizon scales;
as is shown in~(\ref{PBEt1}), the magnetic power spectrum 
for super-horizon modes scales as $\mathcal{P}_B^{1/2} \propto k^{3-s}$, 
and the electric fields have a redder spectrum with $\mathcal{P}_E^{1/2}
\propto k^{2-s}$. 
On the other hand for sub-horizon modes, the magnetic and electric
spectra are identical, scaling as $\mathcal{P}^{1/2} \propto k^2 $.
The wave number where the spectra bends, i.e. $ k \sim a
H_{\mathrm{inf}}$, shifts towards larger $k$ modes as inflation proceeds.

Between the two phases of $I$ evolution, i.e. $a_1 < a < a_2$, the
electromagnetic power spectra conserve their shapes;
the uplifted spectra on large scales (small~$k$) connect to the 
blue spectra on small scales at 
$k \sim a_1 H_{\mathrm{inf}}$, cf.~(\ref{PBEt1tend}) and (\ref{PBEtendt2}).
This is seen in Figure~\ref{fig:sp_12} which shows the spectra at 
$a \approx 3 \times 10^8 a_1 $ ($< a_{\mathrm{end}}$). 

Figure~\ref{fig:sp_23} shows the spectra at $a \approx 2 a_2 $ ($< a_3$). 
The post-inflationary magnetogenesis during $a_2 < a < a_3$ 
enhances the electric fields more strongly than the magnetic fields
(see~(\ref{PBEt2t3})), and thus the two spectra no longer overlap.
The magnetic power now has a term that scales as
$\mathcal{P}_B^{1/2} \propto k^3$, and this component
dominates the spectrum at small scales.

When the coupling~$I$ has approached its final constant value~$I_f$,
i.e. $a > a_3$, 
the magnetic spectrum~$\mathcal{P}_B^{1/2}$ for wave modes in the
range $a_1 H_{\mathrm{inf}} \ll k \ll aH$ 
scales as $k^3$, while 
it remains to be~$ \propto k^{3-s}$ at larger scales $k \ll a_1
H_{\mathrm{inf}}$, cf.~(\ref{PBEt3treh}).
The spectra right before reheating~$a = a_{\mathrm{reh}}$ is shown 
in Figure~\ref{fig:sp_reh}.
The $k$ modes close to the right edge of the plot have already
re-entered the Hubble horizon, and thus the power spectra exhibit
oscillatory behaviors.

\begin{figure}[htbp]
\centering
\begin{minipage}[b]{0.48\linewidth}
  \includegraphics[width=\linewidth]{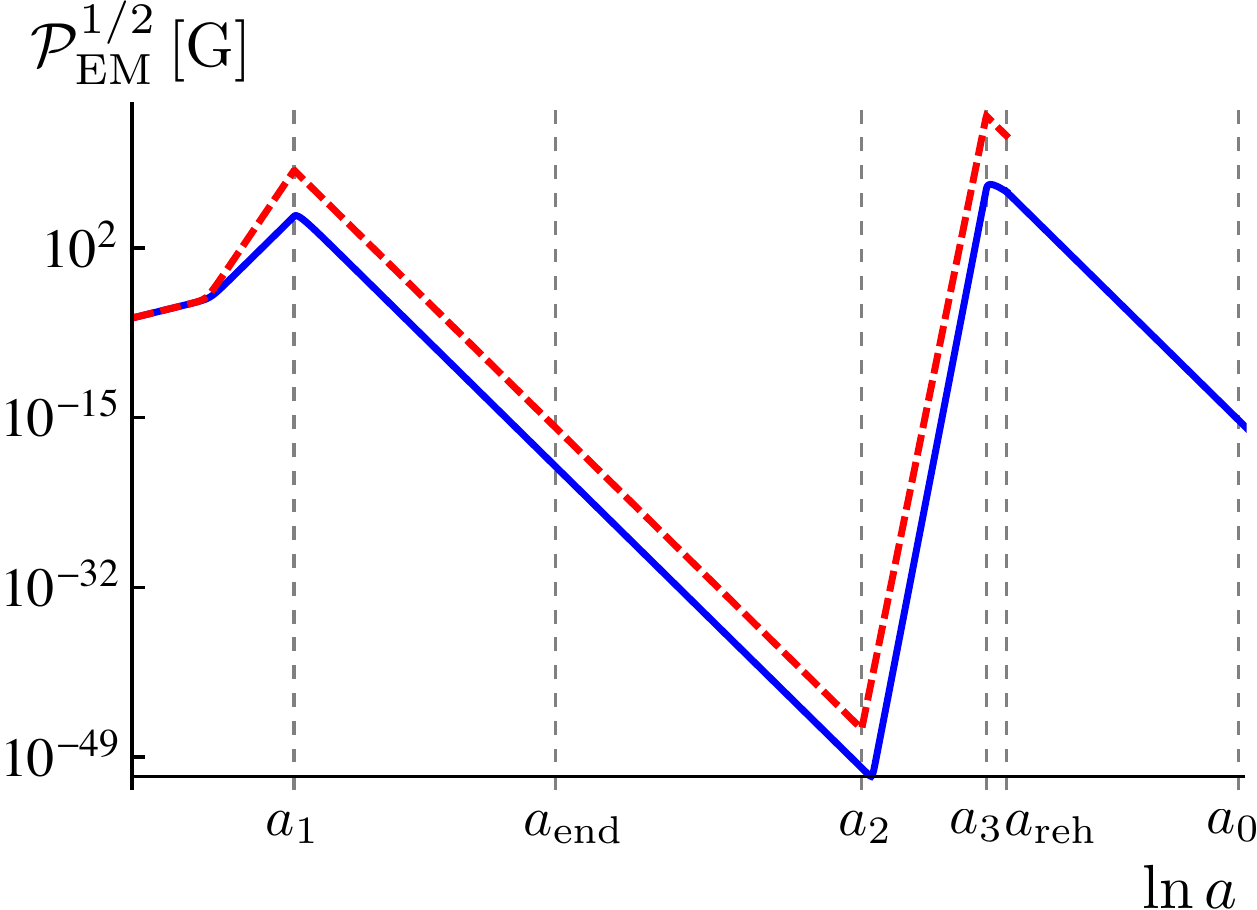}
  \caption{Time evolution of the electromagnetic power spectra for
   $k/a_0 = 1\, \mathrm{Mpc}^{-1}$. $\mathcal{P}_B^{1/2}$: blue solid, $\mathcal{P}_E^{1/2}$: red dashed.}
  \label{fig:PB-evolution}
\vspace{0.35cm}
\end{minipage}
\quad
\begin{minipage}[b]{0.48\linewidth}
\vspace{\baselineskip}
  \includegraphics[width=\linewidth]{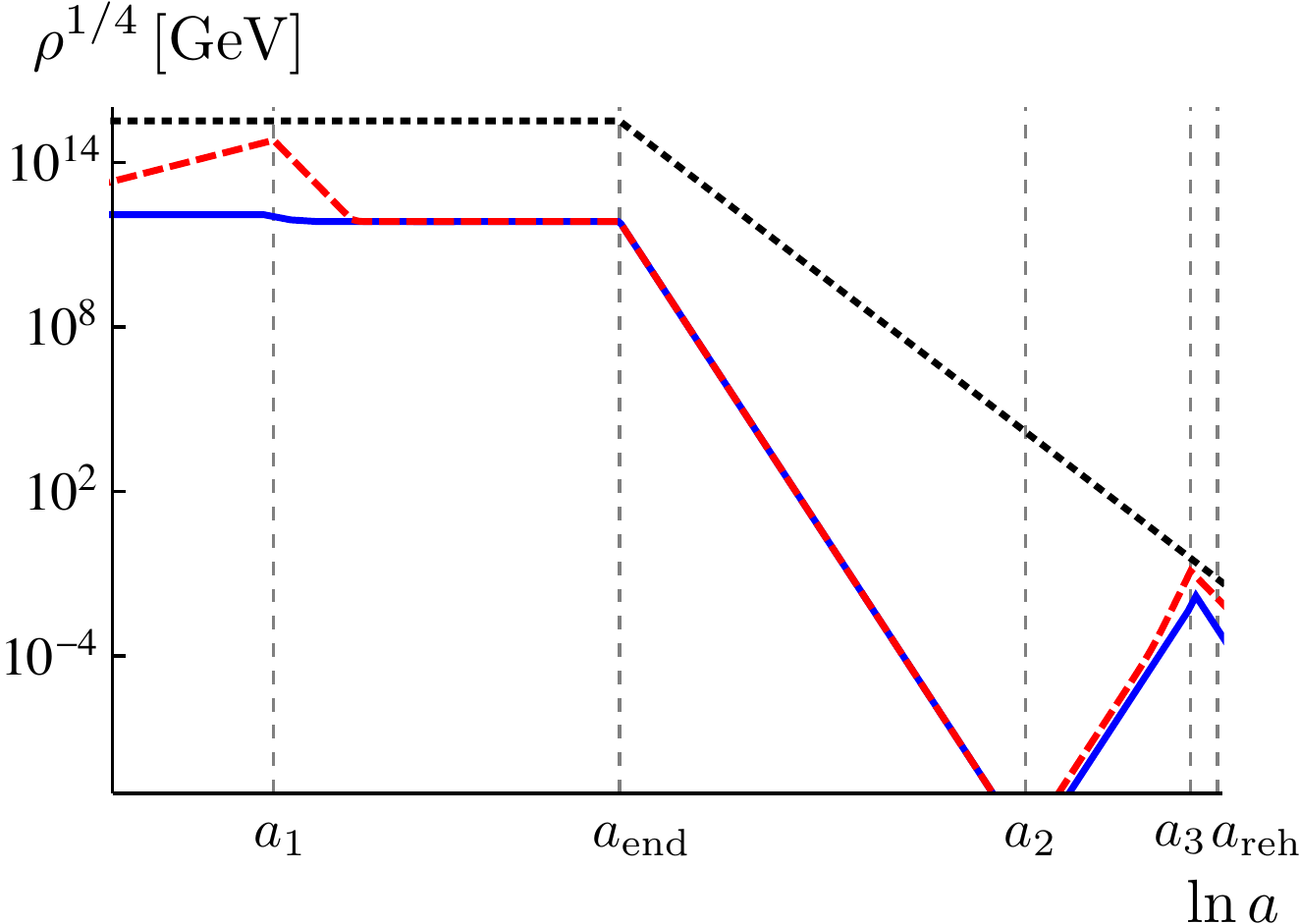}
  \caption{Time evolution of energy densities. Magnetic energy density: blue solid, electric density: red dashed, background density: black dotted.}
  \label{fig:rho-evolution}
\end{minipage}
\end{figure}

\vspace{\baselineskip}

We also plot the electromagnetic power as
a function of~$\ln a$ in Figure~\ref{fig:PB-evolution}, focusing on the
scale $k/a_0 = 1\, \mathrm{Mpc}^{-1}$. 
Note that this wave mode exits the Hubble horizon during inflationary magnetogenesis
(i.e. while $a < a_1$), and re-enters the horizon after
reheating is completed.
During the first phase of magnetogenesis, the magnetic field initially
increases as $B \propto a^{s-2}$, then after the wave mode exits the
horizon as $a^{2s-3}$.
While the coupling~$I$ stays constant the magnetic
amplitude redshifts as $a^{-2}$, but soon after $I$ again becomes time
dependent, $B$ starts to evolve as $a^{2n-3/2}$.
After the standard Maxwell theory is recovered at $a = a_3$, the
magnetic field redshifts as $B \propto a^{-3/2}$, and after reheating as
$a^{-2}$.
For the chosen set of parameters, the generated magnetic field in the
present universe has an amplitude of $B \sim 10^{-15}\, 
\mathrm{G}$ on $\mathrm{Mpc}$ scales.
The evolution of the electric fields is similar to that of the magnetic
fields, but with larger amplitude during most of the expansion history.
The electric power spectrum is not shown beyond $a = a_{\mathrm{reh}}$,
as the conductivity of the universe becomes high during reheating and
the electric fields vanish.

In Figure~\ref{fig:rho-evolution}, the time evolution of the
electromagnetic energy density is shown up until reheating.
The blue solid line denotes the magnetic contribution
(see~(\ref{rho_EM})), and the red dashed line is from the 
electric fields. The two lines lie on top of each other in the time
between the two periods of magnetogenesis. 
We have also plotted the background density~$3 M_p^2
H^2$ shown as the black dotted line. 
As we have discussed in Section~\ref{subsec:energy_bounds}, the energy
density ratio~$\rho_{\mathrm{EM}} / 3 M_p^2 H^2$ peaks at $a = a_1$ and
$a = a_3$, taking values of order $10^{-3}$ and $10^{-2}$ respectively,
under the parameter set chosen in this section.
In the figure, the electric density appears to come close to the background density, 
especially at $a = a_3$. However it should be noted that 
we have plotted $\rho^{1/4}$, and that the units of the vertical axis
vary by many orders of magnitude.

\vspace{\baselineskip}

Thus we have seen that the combined inflationary/post-inflationary
magnetogenesis can efficiently produce large-scale magnetic fields,
as the two phases of magnetogenesis help each other to
avoid the electromagnetic density from dominating the universe.
The post-inflation magnetic enhancement 
relaxes inflationary magnetogenesis, therefore ameliorates 
the electric backreaction problem during inflation.
At the same time,
inflationary magnetogenesis can lift the
magnetic spectrum at large scales (small~$k$), and therefore can reduce the
overall electromagnetic spectrum while maintaining magnetic power at
large scales. 
This feature helps the post-inflation magnetogenesis which alone
would produce too large electromagnetic power at small scales.
On the other hand, the two step magnetogenesis produces large-scale
electric power 
which may dominate the post-inflationary universe, and this has lead to a
new constraint~(\ref{eq5.5}). 
However we have seen through the example case studied in this section 
that such electric densities can be put under control.

As for the dynamical coupling~$I$, it either monotonically decreases in time
or stays constant until it approaches the final value $I_f = 1$.
Therefore the interactions between the electromagnetic fields and other
sectors such as charged fermions are strongly suppressed through most of
the history of the universe.

\section{Examples of Scalar Couplings}
\label{sec:couplings}

Let us show some examples of scalar field couplings that 
realize the time dependent~$I$ considered in the previous sections.
We individually discuss the dynamical couplings during and after
inflation, but they can be combined to drive the two step magnetogenesis.
We focus on couplings with scalar fields, however it should be noted
that non-minimal couplings to gravity could also play a similar role.

Some of the examples in the following are somewhat ad hoc as they are
designed to reproduce the simplified behaviors
for~$I$ in (\ref{Iinf}) and (\ref{Imd}). 
It would be interesting to start from simple forms of scalar
couplings and then discuss magnetogenesis that follows.

We should also remark that in explicitly constructed models, the
produced electromagnetic fields can backreact on the coupled scalar. 
Furthermore, the electromagnetic fields may decay into the scalar particles.
To know whether such effects become important requires a full
treatment of the combined scalar/electromagnetic dynamics. 
We leave this for future work.

\subsection{Couplings During Inflation}
\label{subsec:c_during}

The dynamical coupling that behaves as~(\ref{Iinf}) during inflation can
be realized by the inflaton field, or some other spectator field.

\subsubsection{Inflaton Field}

Let us consider an inflaton field~$\phi$ with a quadratic potential
\begin{equation}
 V(\phi) = \frac{1}{2} m^2 \phi^2.
\end{equation}
Then one can solve the slow-roll approximations $3 H \dot{\phi} \simeq
-m^2 \phi$, $3 M_p^2 H^2 \simeq m^2 \phi^2/2$, and obtain
\begin{equation}
\phi^2 - \phi_1^2 \simeq -4 M_p^2 \ln \frac{a}{a_1}.
\end{equation}
Here $a = a_1$ corresponds to the time when the inflaton field~$\phi$
crosses~$\pm \phi_1$.
As was discussed in previous works such as~\cite{Martin:2007ue}, 
by considering an exponential coupling to the electromagnetic
kinetic term of the form
\begin{equation}
 I(\phi) = I_1 \left\{ 
1 + \exp \left(\frac{s}{4} \frac{\phi^2 - \phi_1^2}{M_p^2}\right)
\right\},
\end{equation}
one sees that it evolves as
\begin{equation}
 I \simeq I_1 \left\{ 1 + \left(\frac{a_1}{a}\right)^s  \right\}.
\label{IasIinf}
\end{equation}
If $s > 0$, then the coupling~$I$ scales as $a^{-s}$ when $\phi^2 >
\phi_1^2$, and then asymptotes to the constant value~$I_1$ for $\phi^2 < \phi_1^2$.
Therefore (\ref{Iinf}) is realized.
The discussion here can be generalized to arbitrary power-law inflaton 
potentials~\cite{Martin:2007ue}.

\subsubsection{Spectator Field}

One can also imagine a case where the electromagnetic fields are coupled
to a scalar that has minimal effect on the inflationary
background. 
Such a spectator field~$\sigma$ can also source the dynamical coupling.
Suppose that $\sigma$ has a quadratic potential
\begin{equation}
 V(\sigma) = \frac{1}{2} m^2 \sigma^2,
\end{equation}
with a mass~$m$ much lighter than the inflationary
Hubble~$H_{\mathrm{inf}}$.
Then the slow-roll equation $3 H_{\mathrm{inf}} \dot{\sigma} \simeq -m^2
\sigma$ is solved as
\begin{equation}
 \sigma \propto a^{-m^2/3 H_{\mathrm{inf}}^2}.
\end{equation}
In this case, a polynomial coupling of the form
\begin{equation}
 I(\sigma) = I_1 \left\{ 
1 + \left(\frac{\sigma }{\sigma_1}\right)^{3s H_{\mathrm{inf}}^2/m^2}
 \right\}
\label{eq6.5}
\end{equation}
evolves as~(\ref{IasIinf}).
However we should also remark that in this example,
$s$ of order unity requires the exponent $3s H_{\mathrm{inf}}^2/m^2$ in
(\ref{eq6.5}) is to be much larger than unity.

\subsection{Couplings After Inflation}
\label{subsec:c_after}

The coupling~(\ref{Imd}) which was discussed for post-inflationary
magnetogenesis can be realized by rolling scalars, or oscillating scalars.

\subsubsection{Rolling Scalars}

Let us again consider a spectator scalar field~$\sigma$ with a quadratic potential
\begin{equation}
 V(\sigma) = \frac{1}{2} m^2 \sigma^2.
\label{Vsigma}
\end{equation}
We suppose that the background universe is effectively matter-dominated, and that
the energy density of~$\sigma$ is tiny 
compared to the total energy of the universe.
If the field's mass is lighter than the Hubble parameter, 
i.e. $m^2 \ll H^2$, then a homogeneous~$\sigma$ field 
follows an attractor solution~\cite{Chiba:2009sj,Kawasaki:2011pd}
\begin{equation}
 \frac{9}{2} H \dot{\sigma} \simeq -m^2 \sigma.
\label{9halves}
\end{equation}
Note that this is similar to the slow-roll approximation during
inflation, except for that the numerical factor in the left hand side is
$9/2$ in an MD universe.
(\ref{9halves}) can be solved using $H^2 \propto a^{-3}$, giving 
\begin{equation}
 \ln \left(\frac{\sigma}{\sigma_2}\right)
 \simeq 
 \frac{2}{27} \frac{m^2}{H_2^2}
\left\{ 1 - \left( \frac{a}{a_2} \right)^3 \right\},
\end{equation}
where the subscript~$2$ denotes quantities measured at a certain time.
Therefore, considering a coupling of the form
\begin{equation}
 I(\sigma) = I_f \left[
1 + 
\left\{
 \ln \left(\frac{\sigma_3}{\sigma_2}\right)^2
\right\}^{n/3}
\left\{
 \ln \left( \frac{\sigma }{\sigma_2} \right)^2
-
\frac{8}{27}\frac{m^2}{H_2^2}
\right\}^{-n/3}
\right],
\end{equation}
where the subscript~$3$ is for quantities at some time after~$t_2$, 
then one sees that the coupling is approximated by
\begin{equation}
 I \simeq I_f
\left[
1 + 
\left\{ \left(\frac{a_3}{a_2}\right)^3 - 1 \right\}^{n/3}
\left\{ \left(\frac{a}{a_2}\right)^3 + 1 \right\}^{-n/3}
\right].
\end{equation}
This coupling reproduces the behavior of (\ref{Imd}) for $n > 0$.

\subsubsection{Oscillating Scalars}

A scalar~$\sigma$ with a quadratic potential~(\ref{Vsigma}) eventually
starts to oscillate when the Hubble parameter becomes comparable to its mass.
The harmonically oscillating field acts like pressureless dust, 
and so the field's oscillation amplitude~$\widetilde{\sigma}$ decays as
$\propto a^{-3/2}$.
Such an oscillating field can also source the dynamical coupling.

Let us consider a coupling as follows,
\begin{equation}
 I(\sigma) = I_f 
\left( 1 + \left| \frac{\sigma}{\widetilde{\sigma}_3} \right|^{2n/3}
\right),
\label{Isigma6.13}
\end{equation}
where we suppose the constants $n$ and $\widetilde{\sigma}_3$ to be positive,
and that $\abs{\sigma}$ initially takes a
field value much larger than $\widetilde{\sigma}_3$.
Prior to the $\sigma$ oscillation, i.e. $ H \gg m$, the coupling~$I$
only slowly evolves. But once the field starts to oscillate,
$I$ decays as $\propto a^{-n}$ when averaging over the oscillations.
As the oscillation amplitude~$\widetilde{\sigma}$ becomes smaller than
$\widetilde{\sigma}_3 $, the coupling $I$ asymptotes to a constant 
value~$I_f$. 
This appears to
reproduce the time dependent behavior~(\ref{Imd}) that was discussed for
post-inflationary magnetogenesis. 

However it should be noted that during the $\sigma$ oscillations, the
coupling~(\ref{Isigma6.13}) also oscillates about~$I_f$. 
Each time $\sigma$ crosses its origin $\sigma = 0$, 
the change in the effective frequency of the mode function~$u_k$ may
become nonadiabatic (see~(\ref{maru1})), 
leading to resonant amplifications of the electromagnetic fields. 
Investigation of such effects goes beyond the scope of this paper,
but it would be interesting to explore the possibility of producing
magnetic fields from parametric resonance.\footnote{However we also note
that, even if resonant amplifications of magnetic fields happen during
inflation, it would not improve much the
situation for inflationary magnetogenesis alone. This is because a rather
strict upper bound on the inflation scale can still be obtained 
independently of how the mode function evolves during
inflation, see~\cite{Fujita:2012rb}.}

A model without parametric resonance may be realized by having multiple
oscillating fields. For example, considering a complex scalar
field~$\Phi$ possessing a potential along the phase direction such as
\begin{equation}
 V(\Phi) =  m^2 \left| \Phi  \right|^2 + 
\left( \lambda \frac{\Phi^c}{M^{c-4}} + \mathrm{h.c.} \right),
\end{equation}
then $\Phi$ can keep rotating in the complex plane 
without crossing the origin.
With a coupling that depends on the decaying radial component,
\begin{equation}
 I(\Phi ) = I_f 
\left( 1 + \left| \frac{\Phi}{\widetilde{\Phi}_3} \right|^{2n/3}
\right),
\end{equation}
the mode function~$u_k$ may evolve adiabatically.

\section{Conclusions}
\label{sec:conc}

In this work, we explored cosmological magnetogenesis during the two
phases when the universe is cold: the inflationary epoch, and the
post-inflationary epoch prior to reheating, during which the universe is
dominated by the oscillating inflaton.
Magnetogenesis in each phase alone are highly
constrained by the strong coupling and backreaction problems, however,
we have found that the combined inflationary {\it and}
post-inflationary magnetogenesis can overcome the difficulties and
efficiently produce large-scale magnetic fields. 
In particular, we demonstrated that the combined
inflationary/post-inflationary magnetogenesis scenario can 
produce magnetic fields stronger than $10^{-15}\, \mathrm{G}$ on Mpc
scales without running into the strong coupling regime, or producing too
large electric fields that would dominate the universe.
The proposed model is compatible even with high scale inflation.

The strong enhancement of the magnetic fields is made possible in the 
two step scenario due to the magnetogenesis in the two
epochs working in very different ways.
The magnetic enhancement in the post-inflationary universe
reduces the need for a significant production of magnetic fields
during inflation, and thus relaxes the constraints on inflationary
magnetogenesis,
including those from the backreaction problem and the excessive
production of cosmological density perturbations.
On the other hand, inflationary magnetogenesis enhances large-scale
magnetic fields and thus can relatively suppress the 
small-scale (but super-horizon) electromagnetic fields 
produced during post-inflationary magnetogenesis. 
Therefore the two phases of magnetogenesis mutually prevent the
electromagnetic fields from dominating the universe in each epoch, 
while maintaining magnetic power at large scales. 
Moreover, we have shown that the net magnetic enhancement from the
two step magnetogenesis is much stronger than 
a naive product of the enhancements from the individual phases of
magnetogenesis.

In order to generate large magnetic fields from the cosmological
background, we considered breaking the conformal invariance of the Maxwell
theory $-I^2 F_{\mu \nu} F^{\mu \nu }/4$ through a time dependent
coupling~$I$ that scales as a power-law of the scale factor. 
As we discussed in Section~\ref{sec:couplings}, such a dynamical~$I$ can
arise from couplings with the inflaton or some other
spectator field(s) that rolls along its effective potential, or
oscillates about the potential minimum. 
Here we note that the produced electromagnetic
fields may backreact on the directly 
coupled scalars in some cases.
To see whether such effects become important requires a detailed
analysis of explicitly constructed models.
We leave this for future work.

The difficulty of post-inflationary magnetogenesis alone was presented
in the form of the strict upper bound on the reheating 
scale $H_{\mathrm{reh}} \ll 10^{-23} \, \mathrm{MeV}$ (\ref{Hrehbound}),
which is incompatible with BBN.
Upon deriving this bound we have assumed the scaling~$I \propto a^{-n}$, 
and so the generality of the reheating bound remains an open question.
For example, electromagnetic couplings with oscillating scalars can lead to further
amplification of the magnetic fields through parametric resonance,
which may allow the post-inflationary universe alone to
create large magnetic fields, without the aid of inflationary magnetogenesis.
Thus it would be interesting to derive a generic bound on post-inflationary
magnetogenesis without specifying how the electromagnetic mode function
evolves in time.
Such an analysis was carried out for inflationary magnetogenesis
in~\cite{Fujita:2012rb}, and it may be possible to apply their
discussions to the post-inflationary universe as well.
We also note that the magnetic enhancement in the post-inflation
universe should be affected by the details of the reheating process.
We have assumed the conductivity of the universe to suddenly become high towards
the end of the inflaton-dominated epoch,
however the conductivity may start to gradually increase
from earlier times if the oscillating inflaton decays perturbatively.
On the other hand, preheating~\cite{Kofman:1994rk,Kofman:1997yn} may
lead to a sudden growth of the conductivity.

Although more detailed work will be required to verify whether there
actually exists intergalactic magnetic fields, investigation of
cosmological magnetic fields may provide an observational window
into the very early universe.
We demonstrated that large-scale magnetic fields can actually be created
from the cosmological background. 
We hope that our mechanism will provide new insights into explaining our
magnetized universe from the cosmological point of view.

\section*{Acknowledgements}

It is a pleasure to thank Niayesh Afshordi, Avery Broderick, Carlos
Palenzuela, Richard Shaw, and Chris Thompson for very useful
discussions. I am also grateful to Shinji Mukohyama for useful discussions 
as well as helpful comments on a draft.





\end{document}